\documentclass[10pt]{article}
\usepackage{xcolor}
\usepackage{fancyhdr}
\usepackage{graphicx}
\usepackage{lineno}
\usepackage{hyperref}
\usepackage{doi}
\usepackage{cite}
\usepackage[top=2cm,left=2cm,right=2cm,top=2cm]{geometry}
\usepackage[english]{babel}
\usepackage[utf8]{inputenc}
\usepackage[T1]{fontenc}
\usepackage{dsfont}
\usepackage{physics}
\usepackage{fontawesome}
\usepackage[]{commath}
\usepackage{amsmath,amsfonts,amssymb}
\usepackage{cancel}
\usepackage{varwidth}
\usepackage[binary-units = true,load-configurations=binary]{siunitx}

\newcommand{\dgr}{\ensuremath{^{\dagger}}}

\begin{document}
% The article title is centered, Large boldface, and should fit in two lines
\begin{center}{\Large \textbf{
Gauss Law, Minimal Coupling and Fermionic PEPS for Lattice Gauge Theories
}}\end{center}

% TODO: write the author list here. Use initials + surname format.
% Separate subsequent authors by a comma, omit comma at the end of the list.
% Mark the corresponding author with a superscript *.
\begin{center}
P. Emonts\textsuperscript{1},\\
E. Zohar\textsuperscript{2},\\
\end{center}

% Format: institute, city, country
\begin{center}
{\bf 1} Max-Planck-Institut f\"ur Quantenoptik, Hans-Kopfermann-Stra\ss e 1, 85748 Garching, Germany\\
{\bf 2} Racah Institute of Physics, The Hebrew University of Jerusalem 91904, Givat Ram, Jerusalem, Israel\\
\end{center}

\begin{center}
\today
\end{center}

% For convenience during refereeing: line numbers
%\linenumbers

\section*{Abstract}
{\bf
In these lecture notes, we review some recent works on Hamiltonian lattice
gauge theories, that involve, in particular, tensor network methods.
The results reviewed here are tailored together in a slightly different
way from the one used in the contexts where they were first introduced.
We look at the Gauss law from two different points of view:
for the gauge field, it is a differential equation, while from
 the matter point of view, on the other hand, it is a simple, explicit algebraic equation.
 We will review and discuss what these two points of view allow and do not allow us to do, in terms of unitarily gauging a pure-matter theory and eliminating the matter from a gauge theory, and relate that to the construction of PEPS (Projected Entangled Pair States) for lattice gauge theories.
}

% Guideline: if your paper is longer that 6 pages, include a TOC
% To remove the TOC, simply cut the following block
\vspace{10pt}
\noindent\rule{\textwidth}{1pt}
\tableofcontents\thispagestyle{fancy}
\noindent\rule{\textwidth}{1pt}
\vspace{10pt}

% Guideline: if your paper is longer that 6 pages, include a TOC
% To remove the TOC, simply cut the following block
%\vspace{10pt}
%\noindent\rule{\textwidth}{1pt}
%\tableofcontents\thispagestyle{fancy}
%\noindent\rule{\textwidth}{1pt}
%\vspace{10pt}

\section{Introduction}

Gauge symmetries, that provide the standard model's description of interactions, are fascinating. 
They have many interesting properties, that make them an excellent playground for endless studies of non-perturbative physics. 
They offer puzzling phenomena to study, understand, and solve, due to their rich symmetry and non-perturbative nature;
perhaps the most famous one is the confinement of quarks in Quantum Chromodynamics (QCD) and other non-Abelian gauge theories~\cite{wilson_confinement_1974}.

Asymptotic freedom~\cite{gross_ultraviolet_1973} tells us that 
the coupling of QCD is small at high energies, allowing one to use perturbation
theory and Feynman diagrams to describe and understand collider physics. 
However, the other, non-perturbative  low energy physics side, is yet to be understood.
A very successful approach to that has been through lattice gauge theories~\cite{wilson_confinement_1974,kogut_introduction_1979}, where spacetime is discretized in a way that allows one to regularize the theory in
a gauge invariant manner, and to perform successful Monte-Carlo
calculations of many important physical properties -- such as the hadronic
spectrum~\cite{flag_working_group_review_2014}. On the other hand, as the Monte-Carlo calculations
are carried out in Euclidean spacetime, they do not allow, in general,
to consider real-time evolution and encounter the fermionic sign
problem~\cite{troyer_computational_2005} in many physically relevant scenarios
required, for example, for the study of exotic phases of QCD~\cite{fukushima_phase_2011}.

Recently, much effort has been put into lattice gauge theories from
the quantum information and computation point of view, suggesting new ways
to tackle these problems. One is quantum simulation~\cite{zohar_quantum_2016,wiese_ultracold_2013,dalmonte_lattice_2016},
which suggests to map the gauge theory degrees of freedom to those
of a controllable quantum system that could be used as an analog experiment,
or a digital quantum computer for gauge theories. 
The other involves the application of tensor network tools, such as MPS or PEPS~\cite{verstraete_matrix_2008,orus_practical_2014}.

Lattice gauge theories, in both methods, are approached in the
Hamiltonian formalism, first introduced by Kogut and Susskind~\cite{kogut_hamiltonian_1975}.
Unlike in path integral methods used widely in particle physics in
general, and in lattice gauge theories in particular, the Hamiltonian
formulation uses states and operators in Hilbert spaces,
that have a very special structure due to the gauge symmetry.

Constructing a quantum simulator or a tensor network state involves
an interesting challenge: one has to reconstruct a physical model
from elementary building blocks. In this process, one faces the most
fundamental elements of a theory, which is decomposed to its smallest
ingredients. This, in some sense, is what we would like to discuss
in this lecture: it will mostly be about enforcing gauge invariance
on quantum states -- the construction of tensor network states for lattice
gauge theories, through a discussion of the Gauss law.

The key ingredient of gauge theories is the local nature of the symmetry:
\emph{a gauge symmetry is a local symmetry}, which means that they involve
many (local) conservation laws, or constraints.
 The gauge symmetry is formulated, in both classical
and quantum gauge theories, by means of the Gauss law, an equation
-- or a set of equations, one per space point -- that relates the gauge
field and the matter fields. Its simplest form, of classical electrodynamics,
is
\begin{equation}
\nabla\cdot\vb{E}\left(\vb{x}\right)=\rho\left(\vb{x}\right)\label{G0}
\end{equation}
-- the divergence of the electric field $\vb{E}$ at each space point $\vb{x}$ equals the density of matter charges there, $\rho$. 
It can be made more complicated when the gauge field is non-Abelian,
become an eigenvalue equation for the so-called \emph{physical states} after
quantization, or a difference equation for lattice gauge theories,
but the essence is the same in all those cases: the matter is the
source of electric fields.

The Gauss law is a static equation, even though its ingredients are,
in general, time dependent. It involves no temporal derivatives; as
the generator of a symmetry, it commutes with the Hamiltonian. It
formulates constants of motion, rather than an equation of motion.
Although it can be obtained from the Euler-Lagrange equations for
fields, it is an artifact of the continuous Lagrangian formalism for fields,
that puts space and time on an equal footing.
However, it cannot be obtained through the Hamilton equations, but rather appears
as a constraint accompanied by a Lagrange multiplier once a Legendre
transformation into the Hamiltonian formalism is carried out. In lattice
formulations with continuous time, which do not put time and space
on an equal footing, it cannot be obtained as an equation of motion 
either. As it is a constraint, it means we can try to solve it, and
if we manage to do that, we plug it into the other equations of motion to reduce the number of degrees of freedom.

In this lecture, we will discuss the nature and possibilities of solving
the Gauss law, in two contexts. First, we will look at it as a differential
equation for the gauge field, and see that it tells us that gauging
a free matter theory (minimal coupling) could not be done, in the
most general setting using a local unitary transformation. On the
other hand, we can break the system into pieces
with simple Gauss laws (and we will explain what "simple" means here),
modify globally invariant Hamiltonians or states by gauging
locally and unitarily each piece alone, and then tailor everything
together; this will be done using the Trotter-Suzuki decomposition
for Hamiltonians, and construction of PEPS - Projected Entangled Pair
States. Second, we will discuss the possibility of solving it for
the matter, and using that for the elimination of the matter degrees
of freedom, or, in case they are fermionic -- their replacement by
spins.

The lecture reviews the works~\cite{zohar_fermionic_2015,zohar_building_2016,zohar_projected_2016,zohar_digital_2017,zohar_digital_2017-1,zohar_combining_2018,zohar_eliminating_2018, zohar_removing_2019}.

\section{Gauging Hamiltonians and Quantum States}

\subsection{Minimal Coupling and its Compact Lattice Formulation}
A conventional textbook approach to gauge theories is to consider
first a non-interacting matter field theory, with some global symmetry:
the reader is suggested to naively try to act on a globally invariant
Lagrangian or Hamiltonian with local transformations instead of global
ones, only to quickly find out that the kinetic (derivative) term
is not invariant under them. Then, in order to obtain something with
a local symmetry nevertheless, another degree of freedom -- the gauge
field -- is introduced as a geometric connection, modifying the previously
quadratic, non-interacting terms into interaction terms of the matter
with the field, in a procedure known as \emph{minimal coupling}. At
this point the gauge field has no dynamics; this could be introduced
by including additional terms in the Hamiltonian or the Lagrangian.

The most famous example, perhaps, is the derivation of the Quantum
Electrodynamics (QED) Lagrangian from that of the free Dirac theory
\cite{peskin_introduction_1995}; one begins with
\begin{equation}
\mathcal{L}_{D}=\overline{\psi}\left(x\right)\left(i\cancel{\partial}-m\right)\psi\left(x\right),
\label{eq:lagrangian_qed_global}
\end{equation}
which is symmetric under the global transformation $\psi\left(x\right)\rightarrow\psi\left(x\right)e^{i\varLambda}$.
The slash notation $\cancel{B}$ is defined as $\cancel{B}\equiv \gamma^\mu B_\mu$, where $\gamma$ are the Dirac gamma matrices, regularly used for describing spin physics in the Dirac theory \cite[Chapter~3]{peskin_introduction_1995}.
In order to make~\eqref{eq:lagrangian_qed_global} symmetric under local transformations of the form
$\psi\left(x\right)\rightarrow\psi\left(x\right)e^{i\varLambda\left(x\right)}$,
one introduces the gauge field $A_{\mu}\left(x\right)$ and replaces
the regular derivative $\partial_{\mu}$ by the covariant one $D_{\mu}=\partial_{\mu}-iA_{\mu}\left(x\right)$,
to obtain instead
\begin{equation}
\widetilde{\mathcal{L}}_{D}=\overline{\psi}\left(x\right)\left(i\cancel{D}-m\right)\psi\left(x\right).
\end{equation}
Finally, the dynamics are obtained by adding the Maxwell part $\mathcal{L}_{EM}=-\frac{1}{4}F_{\mu\nu}\left(x\right)F^{\mu\nu}\left(x\right)$
(where $F_{\mu\nu}\left(x\right) = \partial_{\mu}A_{\nu}\left(x\right) - \partial_{\nu}A_{\mu}\left(x\right)$ is the gauge-invariant field strength tensor).
The complete process could be summarized as
\begin{equation}
\mathcal{L}_{D}\longrightarrow\widetilde{\mathcal{L}}_{D}\longrightarrow\widetilde{\mathcal{L}}_{D}+\mathcal{L}_{EM}.
\end{equation}

One could perform a similar procedure on the lattice. Consider, as
a simple example, a two dimensional spatial lattice, with continuous
time, as in the usual Hamiltonian formulation of lattice gauge theories
\cite{kogut_hamiltonian_1975}. We define a fermionic mode (creation operator)
$\psi^{\dagger}\left(\vb{x}\right)$ at each vertex (lattice site)
$\vb{x}\in\mathbb{Z}^{2}$. We introduce a staggered fermionic
Hamiltonian, where one sublattice corresponds to particles and another
to anti-particles; if one tunes the hopping coefficients in a slightly
different way, which was not done here for the sake of simplicity,
a (doubled) Dirac model is obtained in the continuum limit~\cite{susskind_lattice_1977}:
\begin{equation}
H_{f}=M\underset{\vb{x}}{\sum}\left(-1\right)^{\vb{x}}\psi^{\dagger}\left(\vb{x}\right)\psi\left(\vb{x}\right)+\epsilon\underset{\vb{x},i=1,2}{\sum}\left(\psi^{\dagger}\left(\vb{x}\right)\psi\left(\vb{x}+\vu{e}_{i}\right)+h.c.\right),
\end{equation}
where $\vu{e}_{i}$ is a unit vector in the direction $i=1,2$. $H_f$ is invariant under global phase transformations, generated
by the total fermionic number
\begin{equation}
\mathcal{Q}=\underset{\vb{x}}{\sum}\psi\dgr\left(\vb{x}\right)\psi\left(\vb{x}\right),
\end{equation}
which is a conserved charge.

To make the symmetry local, we introduce on each link $\ell=\left(\vb{x},i\right)$
of the lattice a new Hilbert space for the gauge field.
In the notation $\ell=(\vb{x},i)$, $\vb{x}$ is the lattice position from which the link is emanates to direction $i$.
In a two-dimensional square lattice, $i$ can take the values ${1,2}$.
Along with the Hilbert space, we introduce two conjugate operators: the compact, angular vector potential $\phi\left(\vb{x},i\right)$,
and the electric field $E\left(\vb{x},i\right)$~\cite{kogut_introduction_1979},
\begin{equation}
\left[\phi\left(\vb{x},i\right),E\left(\vb{y},j\right)\right]=i\delta_{\vb{x},\vb{y}}\delta_{ij}.
\end{equation}
$E\left(\vb{x},i\right)$ is a $U(1)$ angular momentum operator, conjugate to the angle $\phi\left(\vb{x},i\right)$; we recognize the Hilbert space on each link of such a compact $U(1)$ lattice gauge theory as this of a particle moving on a ring.

As phase operators are not well defined, we work instead with the
\emph{group element operator}
\begin{equation}
  U\left(\vb{x},i\right)=e^{i\phi\left(\vb{x},i\right)}
\end{equation}
-- playing the role of a Wilson line along the link.

On each link, the Hilbert space may be spanned, in particular, by
two complete sets of states~\cite{zohar_formulation_2015}: either the \emph{magnetic}
one, of group elements (or more accurately, in this case, parameters),
$\left|\phi\right\rangle $, with the orthogonality relation
\begin{equation}
  \braket{\phi'}{\phi}=\delta\left(\phi'-\phi\right)
\end{equation}
or the orthonormal basis of electric field eigenvalues $\ket{\ell}$
(representation basis),
\begin{equation}
  E\ket{\ell}=\ell\ket{\ell},
\end{equation}
for every $\ell\in\mathbb{Z}$: the electric field is a non-bounded operator
with an integer spectrum, and the group element operator serves as
a unitary raising operator of it:
\begin{equation}
  U\ket{\ell}=\ket{\ell+1}.
\end{equation}

In group theory terms, $\ell$ is simply an irreducible representation
of $U\left(1\right)$,
\begin{equation}
\left\langle \phi|\ell\right\rangle =\frac{1}{\sqrt{2\pi}}e^{i\ell\phi}
\end{equation}
- an eigenfunction of the Hamiltonian of a free particle on a ring, $H_{\text{ring}} \propto E^2$.
$U=\int d\phi e^{i\phi}\left|\phi\right\rangle \left\langle \phi\right|$
is a fundamental representation group element operator (and similarly,
one may define for any representation, $U^{k}=\int d\phi e^{ik\phi}\left|\phi\right\rangle \left\langle \phi\right|$,
for which $U^{k}\ket{\ell}=\left|\ell+k\right\rangle $).

Having introduced the gauge field and its Hilbert space, we can minimally
couple it to the matter, by modifying the Hamiltonian to
\begin{equation}
\widetilde{H}_{f}=M\underset{\vb{x}}{\sum}\left(-1\right)^{\vb{x}}\psi^{\dagger}\left(\vb{x}\right)\psi\left(\vb{x}\right)+\epsilon\underset{\vb{x},i=1,2}{\sum}\left(\psi^{\dagger}\left(\vb{x}\right)U\left(\vb{x},i\right)\psi\left(\vb{x}+\vu{e}_{i}\right)+h.c.\right),
\end{equation}
(where $\left(-1\right)^{\vb{x}} = \left(-1\right)^{x_1+x_2}$)
which is gauge invariant, i.e. invariant under local transformations generated by the \emph{Gauss law operators}
\begin{equation}
  \begin{aligned}
    \mathcal{G}\left(\vb{x}\right)&\equiv G\left(\vb{x}\right)-Q\left(\vb{x}\right)\\
    &=\underbrace{E\left(\vb{x},1\right)+E\left(\vb{x},2\right)-E\left(\vb{x}-\vu{e}_{1},1\right)-E\left(\vb{x}-\vu{e}_{2},2\right)}_{G(\vb{x})}-Q\left(\vb{x}\right)\label{G}
  \end{aligned}
\end{equation}
where $G\left(\vb{x}\right)$ is the lattice operator version of the divergence of the electric field at the vertex $\vb{x}$, and the staggered charge is
\begin{equation}
  Q\left(\vb{x}\right)=\psi^{\dagger}\left(\vb{x}\right)\psi\left(\vb{x}\right)-\frac{1}{2}\left(1-\left(-1\right)^{\vb{x}}\right).
\end{equation}

Finally, one can add dynamics to the gauge field, in the form of the
Kogut Susskind (pure gauge) Hamiltonian $H_{KS}=H_{E}+H_{B}$~\cite{kogut_hamiltonian_1975,kogut_introduction_1979},
where
\begin{align}
H_{E}&=\frac{g^{2}}{2}\underset{\vb{x},i}{\sum}E^{2}\left(\vb{x},i\right),\\
H_{B}&=-\frac{1}{g^{2}}\underset{\vb{x},i<j}{\sum}\cos\left(\phi\left(\vb{x},1\right)+\phi\left(\vb{x}+\vu{e}_{1},2\right)-\phi\left(\vb{x}+\vu{e}_{2},1\right)-\phi\left(\vb{x},2\right)\right)
\end{align}
are the electric and magnetic Hamiltonians, respectively, with the coupling constant $g$. 
The complete process can be summarized by
\begin{equation}
  H_{f}\longrightarrow\widetilde{H}_{f}\longrightarrow H=\widetilde{H}_{f}+H_{KS}.
\end{equation}

\subsection{Gauge Symmetry and the Physical Hilbert Space}

The Hilbert space of a lattice gauge theory $\mathcal{H}$, as discussed
above, could be decomposed as
\begin{equation}
  \mathcal{H}\subset\mathcal{H}_{\mathrm{g}}\times\mathcal{H}_{\mathrm{f}},
  \label{HHH}
\end{equation}
where $\mathcal{H}_{\mathrm{g}}$ is the Hilbert space of the gauge
fields on the links, while $\mathcal{H}_{\mathrm{f}}$ represents
the matter degrees of freedom on the vertices. In the fermionic case
it is a Fock space, but one may consider other types of matter as
well. However, not every state in the product space
$\mathcal{H}_{\mathrm{g}}\times\mathcal{H}_{\mathrm{f}}$
 is what we call ``physical'', due to the symmetry and its implications, which is why (\ref{HHH}) is not an equality but rather an embedding.
There have been different approaches to identify the gauge invariant sectors of the Hilbert space within the larger product space, e.g. by finding dualities between gauge invariant systems and spin systems \cite{savit_duality_1980}.
More recently, such an embedding has been discussed in the context of entanglement entropy measurements \cite{buividovich_entanglement_2008}.
In \cite{tagliacozzo_entanglement_2011}, a connection between the Hilbert space structure of gauge theories and tensor networks has been established, by using tensor networks for describing states in the gauge invariant part of the Hilbert space.

The gauge symmetry implies that
\begin{equation}
\left[\mathcal{G}\left(\vb{x}\right),H\right]=0\qquad\forall\vb{x}
\end{equation}
-- every \emph{physical state} $\left|\psi\right\rangle $ is an eigenstate of all the Gauss law operators $\mathcal{G}\left(\vb{x}\right)$, with eigenvalues $q\left(\vb{x}\right)$ which we call \emph{static charges}:
\begin{equation}
  \mathcal{G}\left(\vb{x}\right)\left|\psi\right\rangle =q\left(\vb{x}\right)\left|\psi\right\rangle \quad\Longleftrightarrow\quad G\left(\vb{x}\right)\left|\psi\right\rangle =\left(Q\left(\vb{x}\right)+q\left(\vb{x}\right)\right)\left|\psi\right\rangle.
\end{equation}
This eigenvalue equation is simply the Gauss law -- a quantum lattice
version of the well-known continuum classical equation (\ref{G0}).
The static charge configurations $\left\{ q\left(\vb{x}\right)\right\} $
are constants of motion, splitting the physical Hilbert space into
different superselection sectors,
\begin{equation}
\mathcal{H}=\underset{\left\{ q\left(\vb{x}\right)\right\} }{\cup}\mathcal{H}\left(\left\{ q\left(\vb{x}\right)\right\} \right)
\end{equation}
that are not mixed by the dynamics. When one acts with a local gauge
transformation on a state, a global phase depending on the static
charge configuration sector $\mathcal{H}\left(\left\{ q\left(\vb{x}\right)\right\} \right)$
to which it belongs will appear. Due to the superselection rule we
do not discuss superpositions, and thus only global phases can appear,
and they play no role in quantum mechanics.

Therefore, in some sense, all the states in $\mathcal{H}$ are gauge
invariant -- but with different global phases, depending on the static charges.
In our context, however, a \emph{gauge invariant state} will be such
that is strictly invariant, i.e. without a global phase. This generalizes
also to other, non-Abelian groups, where static charges imply multiplets
of states that are connected through gauge transformation, unless
the static charges are group singlets. From now on, when we discuss
the \emph{physical} or the \emph{gauge invariant Hilbert space}, we
will refer only to the $\mathcal{H}\left(\left\{ q\left(\vb{x}\right)=0\right\} \right)$
sector.

A general physical state $\left|\psi\right\rangle \in\mathcal{H}\left(\left\{ q\left(\vb{x}\right)=0\right\} \right)$
satisfies
\begin{equation}
G\left(\vb{x}\right)\left|\psi\right\rangle =Q\left(\vb{x}\right)\left|\psi\right\rangle
\end{equation}
and thus it may be expressed as a superposition of product states
of gauge field and matter, with the same eigenvalues for $G\left(\vb{x}\right),Q\left(\vb{x}\right)$:
\begin{equation}
  \left|\psi\right\rangle =\underset{\left\{ Q\left(\vb{x}\right)\right\} }{\sum}\alpha\left(\left\{ Q\left(\vb{x}\right)\right\} \right)\left|E\left(\left\{ Q\left(\vb{x}\right)\right\} \right)\right\rangle _{\mathrm{Gauge}} \otimes \left|\left\{ Q\left(\vb{x}\right)\right\} \right\rangle _{\mathrm{Matter}},
\end{equation}
where $\alpha\left(\left\{ Q\left(\vb{x}\right)\right\} \right)\in\mathds{C}$ are the weights for different states in the superposition.
However, note that once $\left|\left\{ Q\left(\vb{x}\right)\right\} \right\rangle _{\mathrm{Matter}}$
is fixed, there is more than one choice for $\left|E\left(\left\{ Q\left(\vb{x}\right)\right\} \right)\right\rangle _{\mathrm{Gauge}}$
(it is, in general, a superposition). Which means that a unitary gauging map of the form
\begin{equation}
\begin{aligned}
&\left(\underset{\left\{ Q\left(\vb{x}\right)\right\} }{\sum}\alpha\left(\left\{ Q\left(\vb{x}\right)\right\} \right)\left|\left\{ Q\left(\vb{x}\right)\right\} \right\rangle _{\mathrm{Matter}}\right)\otimes\left|0\right\rangle _{\mathrm{Gauge}}\\\longrightarrow&\underset{\left\{ Q\left(\vb{x}\right)\right\} }{\sum}\alpha\left(\left\{ Q\left(\vb{x}\right)\right\} \right)\left|E\left(\left\{ Q\left(\vb{x}\right)\right\} \right)\right\rangle _{\mathrm{Gauge}}\left|\left\{ Q\left(\vb{x}\right)\right\} \right\rangle _{\mathrm{Matter}}
\end{aligned}
\end{equation}
(minimal coupling) is, in general, not possible. On the other hand,
once $\left|E\left(\left\{ Q\left(\vb{x}\right)\right\} \right)\right\rangle _{\mathrm{Gauge}}$
is fixed, there could be a unique choice for $\left|\left\{ Q\left(\vb{x}\right)\right\} \right\rangle _{\mathrm{Matter}}$
(depending on the particular model; this is true for our $U\left(1\right)$ staggered case, but not in
general). Therefore a matter eliminating transformation,
\begin{equation}
\begin{aligned}
&\underset{\left\{ Q\left(\vb{x}\right)\right\} }{\sum}\alpha\left(\left\{ Q\left(\vb{x}\right)\right\} \right)\left|E\left(\left\{ Q\left(\vb{x}\right)\right\} \right)\right\rangle _{\mathrm{Gauge}}\left|\left\{ Q\left(\vb{x}\right)\right\} \right\rangle _{\mathrm{Matter}}\\&\longrightarrow\left|0\right\rangle _{\mathrm{Matter}}\otimes\underset{\left\{ Q\left(\vb{x}\right)\right\} }{\sum}\alpha\left(\left\{ Q\left(\vb{x}\right)\right\} \right)\left|E\left(\left\{ Q\left(\vb{x}\right)\right\} \right)\right\rangle _{\mathrm{Gauge}}
\end{aligned}
\end{equation}
exists.

This is strongly related to the Gauss law and its properties -- as we shall discuss below.

\subsection{Can Minimal Coupling be Unitary and Local?}

In quantum mechanics we like unitary transformations. In particular,
when two Hamiltonians are connected by a unitary transformation (just
like any other pair of observables), they will have the same
spectrum.

Let us focus on the first step, $H_{f}\longrightarrow\widetilde{H}_{f}$.
It is clear that $\left[\widetilde{H}_{f},U\left(\vb{x},i\right)\right]=0$
as no gauge field dynamics is introduced at this point. Could we then
find a unitary gauging, or minimal coupling transformation $\mathcal{U}_{G}$
-- for which $\mathcal{U}_{G}H_{f}\mathcal{U}_{G}^{\dagger}=\widetilde{H}_{f}$?
To make the situation even simpler, let us first only deal with a
single link, with two fermionic modes $\psi,\chi$ on its edges. The
globally invariant Hamiltonian is
\begin{equation}
H_{2}=M\left(\psi^{\dagger}\psi-\chi^{\dagger}\chi\right)+\epsilon\left(\psi^{\dagger}\chi+\chi^{\dagger}\psi\right)\label{H2}
\end{equation}
with a global symmetry generated by
\begin{equation}
Q_{2}=\psi^{\dagger}\psi+\chi^{\dagger}\chi.
\end{equation}
Gauging it will lead to
\begin{equation}
  \begin{aligned}
    \widetilde{H}_{2}&=M\left(\psi^{\dagger}\psi-\chi^{\dagger}\chi\right)+\epsilon\left(\psi^{\dagger}U\chi+\chi^{\dagger}U^{\dagger}\psi\right)\\
    &=M\left(\psi^{\dagger}\psi-\chi^{\dagger}\chi\right)+\epsilon\left(\psi^{\dagger}e^{i\phi}\chi+\chi^{\dagger}e^{-i\phi}\psi\right)
  \end{aligned}
\end{equation}
with the symmetry charges (Gauss law operators)
\begin{equation}
\mathcal{G}_{\psi}=E-\psi^{\dagger}\psi\quad;\quad\mathcal{G}_{\chi}=-E-\left(\chi^{\dagger}\chi-1\right)\label{G2}
\end{equation}

Our initial Hilbert space is $\mathcal{H}_{\mathrm{f}}$, a subspace
of the fermionic Hilbert space, that contains the globally invariant
fermionic states. Gauging maps it to $\mathcal{H}_{\mathrm{phys}}$,
a subspace of the product space $\mathcal{H}_{\mathrm{f}}\times\mathcal{H}_{\mathrm{g}}$
(where $\mathcal{H}_{\mathrm{g}}$ is the link's Hilbert space), which
is invariant under the transformations generated by $\mathcal{G}_{\psi}$ and $\mathcal{\mathcal{G}_{\chi}}$.
We will replace the initial space by $\mathcal{H}_{\mathrm{f}}\times\left\{ \left|0\right\rangle \right\} $,
where $E\left|0\right\rangle =0$. We would like to construct a unitary
transformation $\mathcal{U}_{G}$, that maps $\mathcal{H}_{\mathrm{f}}\times\left\{ \left|0\right\rangle \right\} $
to $\mathcal{H}_{\mathrm{phys}}$ -- that is, it will entangle the vertices and link degrees of freedom, in a way that respects
the symmetry generated by $\mathcal{G}_{\psi}$ and $\mathcal{\mathcal{G}_{\chi}}$.

We will have to act on the state $\ket{0}$ in a way
that will introduce the desired conservation laws. Since they have
to be consistent with each other, we need some relation between the
two fermionic number operators, but this is satisfied by our initial,
globally invariant elements of $\mathcal{H}_{\mathrm{f}}$. So we
simply have to identify $E$ with $\psi\dgr\psi$ or $\chi\dgr\chi-1$: that is, to raise the electric field on the link by an amount $\psi\dgr\psi$,
or lower it by $\chi\dgr\chi-1$. Let us choose the first option.
This is done by the unitary~\cite{zohar_combining_2018}
\begin{equation}
  \mathcal{U}_{G}=\int d\phi e^{i\phi\psi^{\dagger}\psi}\dyad{\phi},
\end{equation}
where $\left|\phi\right\rangle $ is a phase eigenstate, such that
\begin{equation}
  U=\int d\phi e^{i\phi}\dyad{\phi}.
\end{equation}

It is a controlled operation that could be seen either as raising
the electric field $E$ with respect to the fermionic charge $\psi^{\dagger}\psi$, or as rotating
the fermionic operators $\psi^{\dagger},\psi$ with respect to the phase operator $\phi$. While the first interpretation suits the point of view of the quantum state, the second one
suits that of the Hamiltonian, explaining us intuitively
that, indeed,
\begin{equation}
\mathcal{U}_{G}H_{2}\mathcal{U}_{G}^{\dagger}=\mathcal{U}_{G}\left(H_{2}\otimes\vb{1}_{\mathrm{link}}\right)\mathcal{U}_{G}^{\dagger}=\widetilde{H}_{2}
\end{equation}
-- so we see, that unitary gauging was obtained by solving the simple
Gauss law~\eqref{G2} which was both local and uniquely solvable.

Could we now extend this to the Hamiltonian $H_{f}$ in arbitrary
dimensions? The answer is no. Of course, for each link $\ell=\left(\vb{x},i\right)$
we can define a gauging transformation
\begin{equation}
  \mathcal{U}_{G}\left(\ell\right)=\mathcal{U}_{G}\left(\vb{x},i\right)=\int d\phi e^{i\phi\psi^{\dagger}\left(\vb{x}\right)\psi\left(\vb{x}\right)}\left|\phi\right\rangle \left\langle \phi\right|_{\ell}
\end{equation}
such that
\begin{equation}
  \mathcal{U}_{G}\left(\vb{x},i\right)\psi^{\dagger}\left(\vb{x}\right)\psi\left(\vb{x}+\vu{e}_{i}\right)\mathcal{U}_{G}^{\dagger}\left(\vb{x},i\right)=\psi^{\dagger}\left(\vb{x}\right)U\left(\vb{x},i\right)\psi\left(\vb{x}+\vu{e}_{i}\right).
\end{equation}
However, each mode appears in more than one link, and applying the
product $\underset{\ell}{\prod}\mathcal{U}_{G}\left(\ell\right)$
to the whole Hamiltonian $H_{f}$ will give rise to a very complicated
and messy expression which is not $\widetilde{H}_{f}$: since now
each Gauss law involves more than one electric field, all fermionic
operators will be rotated with respect to the gauge fields on all
the links around them.

Let us move on more slowly then, and extend our one link system to
an open line. The globally invariant Hamiltonian is
\begin{equation}
H_{1d}=M\underset{x}{\sum}\left(-1\right)^{x}\psi^{\dagger}\left(x\right)\psi\left(x\right)+\epsilon\left(\psi^{\dagger}\left(x\right)\psi\left(x+1\right)+h.c.\right)\label{H2-1}
\end{equation}
with a global symmetry generated by
\begin{equation}
Q_{1d}=\underset{x}{\sum}\psi^{\dagger}\left(x\right)\psi\left(x\right)
\end{equation}
Gauging it will lead to
\begin{align}
\widetilde{H}_{1d}&=M\underset{x}{\sum}\left(-1\right)^{x}\psi^{\dagger}\left(x\right)\psi\left(x\right)+\epsilon\left(\psi^{\dagger}\left(x\right)U\left(x\right)\psi\left(x+1\right)+h.c.\right)\\
&=M\underset{x}{\sum}\left(-1\right)^{x}\psi^{\dagger}\left(x\right)\psi\left(x\right)+\epsilon\left(\psi^{\dagger}\left(x\right)e^{i\phi\left(x\right)}\psi\left(x+1\right)+h.c.\right)
\end{align}
with the symmetry charges (Gauss law operators)
\begin{equation}
\mathcal{G}\left(x\right)=\underbrace{E\left(x\right)-E\left(x-1\right)}_{G(x)}-Q\left(x\right)\label{G2-1}
\end{equation}

We can solve for the electric field in a non-local way, e.g. by
\begin{equation}
E\left(x\right)=\underset{y<x}{\sum}Q\left(y\right).
\end{equation}
If we like to repeat the single link procedure, we will need to
put initially a trivial $\left|0\right\rangle $ state on each link,
and then change it by an amount of $\underset{y<x}{\sum}Q\left(y\right)$.
This leads to a nonlocal gauging transformation, of the form
\begin{equation}
  \mathcal{U}_{G}^{\left(1\right)}=\int\left(\underset{x}{\prod}d\phi_{x}\right)e^{-i\underset{x}{\sum}\left(\overset{x-1}{\underset{y=0}{\sum}}\phi\left(y\right)\right)\psi^{\dagger}\left(x\right)\psi\left(x\right)}\left|\left\{ \phi\right\} \right\rangle \left\langle \left\{ \phi\right\} \right|
\end{equation}
which gives us what we want. Its inverse form can be used for eliminating
the gauge field degrees of freedom in the one dimensional case, when
one starts with the gauge invariant theory, but it will introduce
some nonlocal interactions (from $H_E$). This was used both for MPS computations~\cite{banuls_mass_2013} and quantum simulation~\cite{martinez_real-time_2016}.

We have seen that in the case of a single link we can gauge with a unitary
-- a controlled operation -- because the Gauss law has a unique solution.
For a one dimensional system, we can gauge by a unitary transformation
again, since a unique solution exists, but we lose locality. In more
dimensions we cannot do even that anymore.

In order to understand it better, and in a more general way, let us
look again at the equation that is responsible for all that -- the
Gauss law. Either its lattice form~\eqref{G}, or the continuum form,
\begin{equation}
  \nabla\cdot\vb{E}\left(\vb{x}\right)=\rho\left(\vb{x}\right)\label{Gc}.
\end{equation}

This is the equation that manifests the local symmetry, introduced
to the system in the minimal coupling procedure, relating the gauge
(electric) field to the matter, that existed before. In the process
of gauging, we ``complete'' a physical setting that only had the
matter field to one that involves the gauge field too.

Suppose, we wish to gauge a globally invariant Hamiltonian, or a state,
by introducing a gauge field that will satisfy~\eqref{Gc}. For that, we have to solve the equation for the electric field. 
However, such a solution is, in general, not unique:~\eqref{Gc} is a single, non-vector differential (or difference, on the lattice) equation, for a vector field. 
A general gauge invariant state could be a superposition of many different states satisfying the Gauss law, even if the matter charges are fixed. 
If we have no unique way to determine the gauge field configuration from the matter, a unitary transformation cannot be used for gauging.

We see, therefore, that gauging by a local unitary will generally not work:
Minimally coupling is generally not a unitary modification of the system.
The only scenarios that allow one to gauge using a unitary and local transformation are those in which the Hamiltonian or the
state to be gauged are constructed out of separate, decoupled local parts, each of which could be separately gauged in a unique way. 
Only after gauging, one tailors the already gauged pieces together, and introduces nonlocal correlations. 
That is, we do not transform the whole object (Hamiltonian / state), but rather its building blocks, before connecting them. 

\subsection{The Case of Hamiltonians: Trotterized Gauge Invariant Dynamics}

From a Hamiltonian point of view, unitary gauging of separate building blocks can be achieved using trotterized time evolution. 
The Trotter-Suzuki decomposition~\cite{suzuki_decomposition_1985}, which is very useful for quantum simulation~\cite{lloyd_universal_1996}, allows
one to approximate the time evolution of a Hamiltonian by a product
of Trotter steps: short time evolutions of different parts of the
Hamiltonian, that do not necessarily commute. It is exact if each
Trotter step\emph{ }is infinitesimally short; otherwise, for finitely
short steps, error bounds can be computed.

For our purposes, we will break the Hamiltonians into separate link
contributions, $\widetilde{H}_{f}=\underset{\ell}{\sum}\widetilde{H}_{f,\ell}$
and $H_{f}=\underset{\ell}{\sum}H_{f,\ell}$. This decomposition allows
us to gauge separately:
\begin{equation}
\mathcal{U}_{G}\left(\ell\right)H_{f,\ell}\mathcal{U}_{G}^{\dagger}\left(\ell\right)=\widetilde{H}_{f,\ell}
\end{equation}
such that
\begin{equation}
e^{-i\widetilde{H}_{f}t}=\underset{N\rightarrow\infty}{\mathrm{lim}}\left[\underset{\ell}{\prod}e^{-i\widetilde{H}_{\ell}t/N}\right]^{N}=\underset{N\rightarrow\infty}{\mathrm{lim}}\left[\underset{\ell}{\prod}\left(\mathcal{U}_{G}\left(\ell\right)e^{-iH_{\ell}t/N}\mathcal{U}_{G}^{\dagger}\left(\ell\right)\right)\right]^{N}.
\end{equation}
Choosing a finite but large $N$ this can be used as an approximate time
evolution, as widely done for quantum simulation purposes in other
contexts,
\begin{equation}
e^{-i\widetilde{H}_{f}t}\approx\left[\underset{\ell}{\prod}\left(\mathcal{U}_{G}\left(\ell\right)e^{-iH_{\ell}t/N}\mathcal{U}_{G}^{\dagger}\left(\ell\right)\right)\right]^{N}.
\end{equation}

Why does that work? Because every Trotter step involves the dynamics of a single link Hamiltonian, just like  $H_{2}$ we considered above~\eqref{H2},
and its gauging gives rise to simple, one dimensional Gauss laws generated
by operators such as those of~\eqref{G2} rather than the ambiguous,
vector Gauss laws of the general case~\eqref{G}. They are only ``switched
on'' once a sequence of Trotter steps is considered, as symmetries
of the complete time evolution.

Some errors due to the trotterization (non-commutativity of intersecting
links contributions to the Hamiltonian) will arise; note, however,
that these errors have nothing to do with the gauge symmetry -- each
evolution step is already gauge invariant and thus, although we obtain
an approximate time evolution, the symmetry is exact. This was used
in~\cite{zohar_digital_2017-1,zohar_digital_2017,bender_digital_2018} for digital quantum simulation of
lattice gauge theories (using a more ``economical'' ways of breaking
the Hamiltonian to less pieces; the important thing is that intersecting
links, involving common fermions, will not be included in the same
Trotter step).

\subsection{The Case of Quantum States: Gauging PEPS }

We can also consider the states' point of view: one can try to construct
the state from some local ingredients, that enable separate, individual
actions of gauging transformations, and then, after gauging, project
all the local pieces together in a way that forms a non-trivial, interacting
state.

This can be done in the process we shall describe now. As we shall
see, the states we construct in this way are nothing but PEPS -- Projected
Entangled Pair States~\cite{verstraete_matrix_2008} -- but with a local symmetry,
following the procedure of~\cite{zohar_building_2016}.

\emph{Step 1.} At each site, construct a state $\left|A\left(\vb{x}\right)\right\rangle $
that will involve the physical matter degree of freedom, and some
virtual degrees of freedom as well. The virtual degrees of freedom
are associated with the edges of links starting or ending at the vertex,
and out of them we will construct the \emph{virtual electric fields},
$E_{0}\left(z\right)$, where $z=r,u,l,d$ denotes the edge: right,
up (outgoing) and left, down (ingoing). The state will be constructed
in a way that a Gauss law will be satisfied by the physical charge
and the virtual electric fields: i.e., we demand that it will be invariant
under transformations generated by
\begin{equation}
\mathcal{G}_{0}\equiv G_{0}-Q=E_{0}\left(r\right)+E_{0}\left(u\right)-E_{0}\left(l\right)-E_{0}\left(d\right)-Q,
\end{equation}
where $Q$ is the charge associated with the physical matter degree
of freedom at the vertex, which does not have to be fermionic. At
each vertex we can define the virtual transformations
\begin{equation}
W_{z}\left(\Lambda\right)=e^{i\varLambda E_0\left(z\right)}
\end{equation}
and the physical one
\begin{equation}
V\left(\Lambda\right)=e^{i\varLambda Q}
\end{equation}
such that
\begin{equation}
e^{i\varLambda\mathcal{G}_{0}}=W_{r}W_{u}W_{l}\dgr W_{d}\dgr V\dgr.
\end{equation}
This allows us to write the symmetry condition of $\left|A\left(\vb{x}\right)\right\rangle $
in the conventional form used in the context of PEPS; at each vertex,
the action with a group transformation on the physical degree of freedom
is equivalent to acting on the virtual ones,

\begin{minipage}{\textwidth}
  \begin{equation}
    V\ket{A} =W_{r}W_{u}W_{l}\dgr W_{d}\dgr \ket{A}.
    \label{eq:peps_invariance}
  \end{equation}
  \begin{center}
    \includegraphics{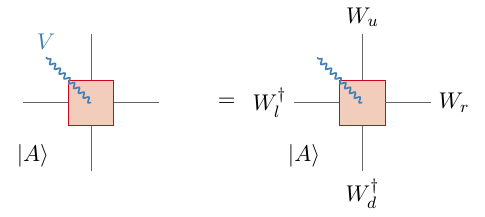}
  \end{center}
\end{minipage}

One nice and useful feature of tensor networks is their pictorial representation.
All manipulations that are carried out with tensor networks can be conveniently depicted.
Thus, the drawing below~\eqref{eq:peps_invariance} yields the same information as the equation itself.
Readers who are not familiar with this notation can find a short introduction in appendix~\ref{app:tn_notation} or in~\cite{bridgeman_hand-waving_2017,orus_practical_2014}.
In these notes, we chose some conventions for the legs to make the understanding of tensor network figures as easy as possible.
Virtual legs are depicted as thin, gray lines. 
The physical legs related to fermions are drawn as thick, blue, wavy lines, while legs related to the local gauging are shown as thick, black, wavy lines.
Projectors are colored in green and fiducial states, i.e. states that are located on the vertices, are shown in red.

\emph{Step 2.} We take a product of the local matter states $\left|A\left(\vb{x}\right)\right\rangle $
with the gauge field states $\left|0\right\rangle $ on all the links
starting at that vertex --
\begin{equation}
\left|A\left(\vb{x}\right)\right\rangle \longrightarrow\left|A\left(\vb{x}\right)\right\rangle \left|0\right\rangle _{\vb{x},1}\left|0\right\rangle _{\vb{x},2}\equiv\left|A\left(\vb{x}\right)\right\rangle \left|0\right\rangle _{\vb{x}}
\end{equation}

\emph{Step 3.} On each $\left|A\left(\vb{x}\right)\right\rangle $ we act
with two gauging transformations $\mathcal{U}_{G}$, rotating (and
entangling) the virtual degrees of freedom corresponding to links
beginning at the vertex with respect to the gauge field degrees of
freedom on that link: $\mathcal{U}_{G}\left(\vb{x},1\right)\mathcal{U}_{G}\left(\vb{x},2\right)$.
The local states obtained by that,
\begin{equation}
\left|A_{G}\left(\vb{x}\right)\right\rangle =\mathcal{U}_{G}\left(\vb{x},1\right)\mathcal{U}_{G}\left(\vb{x},2\right)\left|A\left(\vb{x}\right)\right\rangle \left|0\right\rangle _{\vb{x}}
\end{equation}
will obey two symmetries, connecting the physical fields and the virtual
fields on the outgoing links,
\begin{equation}
V_{r,u}\left|A_{G}\right\rangle =W_{r,u}\left|A_{G}\right\rangle
\end{equation}
where
\begin{equation}
V_{r}\left(\Lambda\right)=e^{i\varLambda E\left(\vb{x},1\right)},\quad V_{u}\left(\Lambda\right)=e^{i\varLambda E\left(\vb{x},2\right)}.
\end{equation}
\begin{center}
  \includegraphics{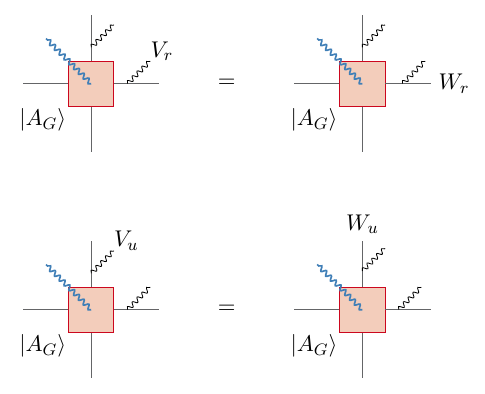}
\end{center}

These are generated by operators such as those we considered in the
one-link case~\eqref{G2}, i.e. connecting each physical electric
field only to one rotated degree of freedom (the virtual one in this
case) and thus gauging is possible. 
As a result, these states will also obey modified Gauss laws, in which the physical electric fields on the outgoing links (up, right) replace the virtual ones, while the virtual fields remain for the ingoing links (down, left):
\begin{equation}
V\left|A_{G}\right\rangle =V_{r}V_{u}W_{l}^{\dagger}W_{d}^{\dagger}\left|A_{G}\right\rangle
\end{equation}
\begin{center}
  \includegraphics{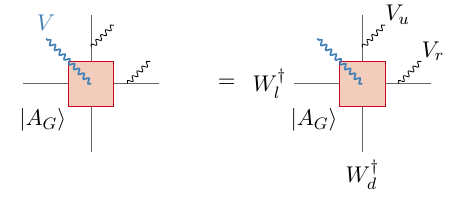}
\end{center}

These Gauss laws are already ``half-way'' to complete physical ones;
we only need to contract the local states to one another, as done
in the next step.

\emph{Step 4.} We project the virtual ingredients into maximally entangled pairs
connecting the edges of links $\left|\omega_{\ell}\right\rangle $,
satisfying the symmetry relations
\begin{align}
W_{r}^{\dagger}\left(\vb{x}\right)W_{l}\left(\vb{x}+\vu{e}_{1}\right)\left|\omega_{\vb{x},1}\right\rangle &=\left|\omega_{\vb{x},1}\right\rangle\\
W_{u}^{\dagger}\left(\vb{x}\right)W_{d}\left(\vb{x}+\vu{e}_{2}\right)\left|\omega_{\vb{x},2}\right\rangle &=\left|\omega_{\vb{x},2}\right\rangle
\end{align}
In the pictorial language of tensor networks, we denote the projector $\omega$ as a green square.
The application of operators is marked by writing their names next to the legs.
\begin{center}
  \includegraphics{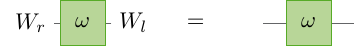}
\end{center}

This will remove all the virtual degrees of freedom, after having
used them for contracting the whole state, which is not a trivial
product state anymore.

The final state
\begin{equation}
\left|\psi\right\rangle =\underset{\text{Step 4}}{\underbrace{\underset{\ell}{\otimes}\left\langle \omega_{\ell}\right|}}\underset{\text{Step 3}}{\underbrace{\underset{\ell}{\prod}\mathcal{U}_{G}\left(\ell\right)}}\underset{\vb{x}}{\otimes}\underset{\text{Step 2}}{\underbrace{\underset{\text{Step 1}}{\underbrace{\left|A\left(\vb{x}\right)\right\rangle}} \left|0\right\rangle _{\vb{x}}}}
\end{equation}
will have the desired physical symmetry -- local gauge invariance.

Steps 1 and 4 can be recognized as the construction of a
PEPS: we project local states with some virtual symmetry to maximally
entangled pair states. In fact, this is a PEPS as well -- a gauged
one. The lesson we learn from that is that in order to obtain a lattice gauge theory PEPS,
we can first write down
a globally invariant PEPS, describing the matter degrees of freedom
of the desired gauge theory,
\begin{equation}
\left|\psi_{0}\right\rangle =\underset{\text{Step 4}}{\underbrace{\underset{\ell}{\otimes}\left\langle \omega_{\ell}\right|}}\underset{\text{Step 1}}{\underbrace{\underset{\vb{x}}{\otimes}\left|A\left(\vb{x}\right)\right\rangle}}
\end{equation}
out of which the gauge invariant PEPS, with dynamical gauge fields,
may be obtained in the process described above, that ``interrupts''
the usual PEPS construction in the form of step 2 and step 3. This is exactly
what we discussed above: gauging, or minimally coupling, by a unitary
transformation, must be done before we contract the local ingredients,
that is, on a product state that allows us to act separately on different
links. Only after gauging we project to a nontrivial physical state
of the whole lattice, at step 4: $\ket{\psi}$ is not
gauged by directly acting with a unitary transformation on $\ket{\psi_{0}}$
; rather, a modification is required before the system's ingredients
are connected.

The nice ``feature'' that we get is that many symmetry properties
of the original, globally invariant PEPS $\left|\psi_{0}\right\rangle $
``survive'' the gauging procedure, such as spatial symmetries (translation,
lattice rotation, lattice inversion etc. -- anything that might have
existed prior to gauging) with the right modification for the gauge
field transformation laws (e.g. translation invariance of $\left|\psi_{0}\right\rangle $
can be made charge conjugation symmetry of $\left|\psi\right\rangle $).
And, of course, the global symmetry becomes local. Therefore, it is
enough to construct a state with a global symmetry, and the one with
local gauge invariance is obtained immediately.

For the sake of completeness, and for readers not familiar with PEPS,
let us show that the PEPS $\ket{\psi_{0}}$ and $\ket{\psi}$
indeed have the right symmetries. First, let us transform $\left|\psi_{0}\right\rangle $
with a global transformation and see what happens:\\
\begin{minipage}{\textwidth}
  \begin{equation}
    \begin{aligned}
      e^{i\varLambda\underset{\vb{x}}{\sum}Q\left(\vb{x}\right)}\left|\psi_{0}\right\rangle  & =  \underset{\ell}{\otimes}\left\langle \omega_{\ell}\right|\underset{\vb{x}}{\otimes}\left(V\left(\vb{x}\right)\left|A\left(\vb{x}\right)\right\rangle \right)\\
      & = \underset{\ell}{\otimes}\left\langle \omega_{\ell}\right|\underset{\vb{x}}{\otimes}\left(W_{r}\left(\vb{x}\right)W_{u}\left(\vb{x}\right)W_{l}^{\dagger}\left(\vb{x}\right)W_{d}^{\dagger}\left(\vb{x}\right)\left|A\left(\vb{x}\right)\right\rangle \right)\\
      & = \underset{\vb{x}}{\otimes}\left(\left\langle \omega_{\vb{x},1}\right|W_{r}\left(\vb{x}\right)W_{l}^{\dagger}\left(\vb{x}+\vu{e}_{1}\right)\left\langle \omega_{\vb{x},2}\right|W_{u}\left(\vb{x}\right)W_{d}^{\dagger}\left(\vb{x}+\vu{e}_{2}\right)\right)\underset{\vb{x}}{\otimes}\left|A\left(\vb{x}\right)\right\rangle \\
      & = \underset{\ell}{\otimes}\ket{\omega_{\ell}}\underset{\vb{x}}{\otimes}\ket{A\left(\vb{x}\right)}\\
      & = \ket{\psi_{0}}
    \end{aligned}
  \end{equation}
  \begin{center}
    \includegraphics[scale=0.45]{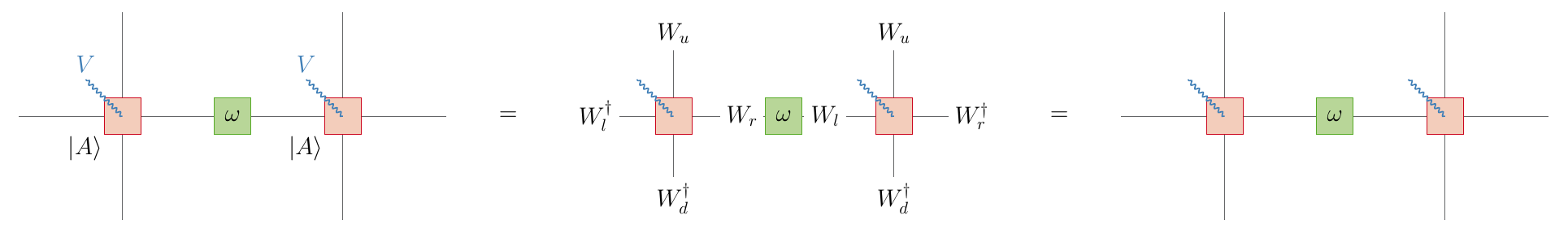}
  \end{center}
\end{minipage}
-- the transition from the first line to the second is thanks to the
symmetry property of $\ket{A\left(\vb{x}\right)}$;
from the third to the fourth, it is the symmetry property of the bond
states $\left\langle \omega_{\ell}\right|$.

In the local case,\\
\begin{minipage}{\textwidth}
\begin{equation}
  \begin{aligned}
    e^{i\varLambda\mathcal{G}\left(\vb{x}_{0}\right)}\ket{\psi}
    &=\underset{\ell}{\otimes}
    \begin{aligned}[t]
      \bra{\omega_{\ell}}V_{r}\left(\vb{x}_{0}\right)V_{u}\left(\vb{x}_{0}\right)V\dgr\left(\vb{x}_{0}\right)&\ket{A_{G} \left(\vb{x}_{0}\right)}\\
      V_{r}\dgr\left(\vb{x}_{0}-\vu{e}_{1}\right)&\ket{A_{G}\left(\vb{x}_{0}-\vu{e}_{1}\right)}\\
      V_{u}\dgr\left(\vb{x}_{0}-\vu{e}_{2}\right)&\ket{A_{G}\left(\vb{x}_{0}-\vu{e}_{2}\right)} 
      \underset{\substack{\vb{x}\neq\vb{x}_{0},\\\vb{x}_{0}-\vu{e}_{1},\\\vb{x}_{0}-\vu{e}_{2}}}{\otimes}\ket{A_{G}\left(\vb{x}\right)}
    \end{aligned}\\
    &=\underset{\ell}{\otimes}
    \begin{aligned}[t]
      \bra{\omega_{\ell}}W_{l}\left(\vb{x}_{0}\right)W_{d}\left(\vb{x}_{0}\right)&\ket{A_{G}\left(\vb{x}_{0}\right)}\\
      W_{r}\dgr\left(\vb{x}_{0}-\vu{e}_{1}\right)&\ket{A_{G}\left(\vb{x}_{0}-\vu{e}_{1}\right)}\\
      W_{u}\dgr\left(\vb{x}_{0}-\vu{e}_{2}\right)&\ket{A_{G}\left(\vb{x}_{0}-\vu{e}_{2}\right)}
      \underset{\substack{\vb{x}\neq\vb{x}_{0},\\\vb{x}_{0}-\vu{e}_{1},\\\vb{x}_{0}-\vu{e}_{2}}}{\otimes}\left|A_{G}\left(\vb{x}\right)\right\rangle
    \end{aligned}\\
    &=\underset{\ell}{\otimes}\bra{\omega_{\ell}}W_{l}\left(\vb{x}_{0}\right)W_{r}\dgr\left(\vb{x}_{0}-\vu{e}_{1}\right)W_{d}\left(\vb{x}_{0}\right)W_{u}\dgr\left(\vb{x}_{0}-\vu{e}_{2}\right)
    \underset{\vb{x}}{\otimes}\ket{A_{G}\left(\vb{x}\right)}\\
    &=\underset{\ell}{\otimes}\bra{\omega_{\ell}}\underset{\vb{x}}{\otimes}\ket{A_{G}\left(\vb{x}\right)}\\
    &=\ket{\psi}
  \end{aligned}
\end{equation}
\begin{center}
    \includegraphics[scale=0.45]{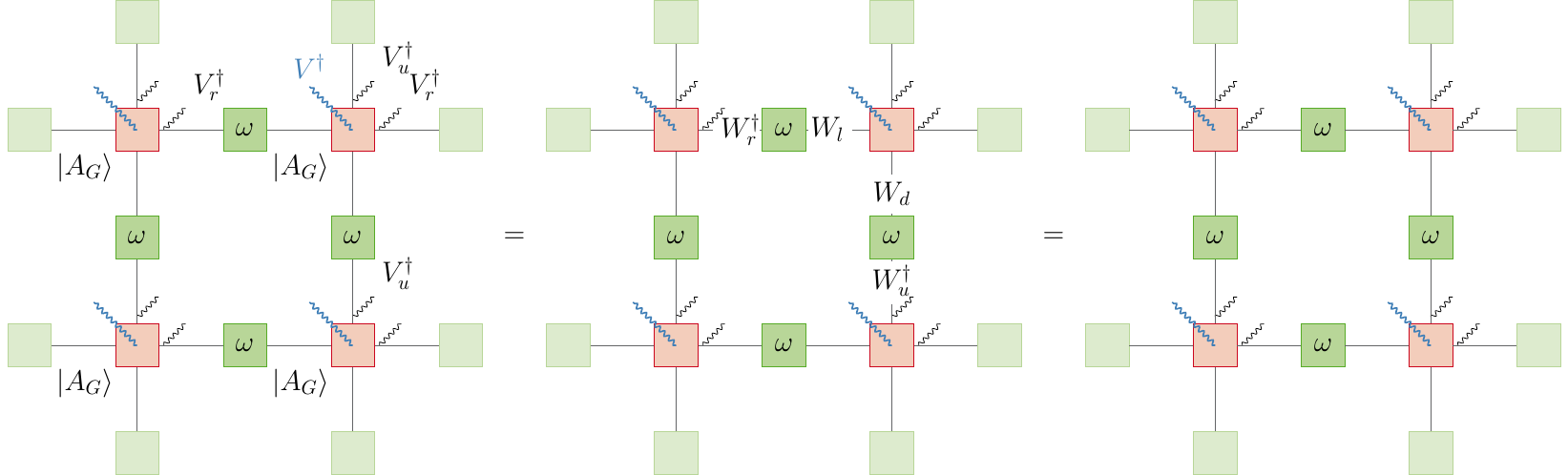}
\end{center}
\end{minipage}

As a final remark, note that this formulation can also be used for
the construction of pure gauge states, with no physical matter: in
that case, the states $\left|A\left(\vb{x}\right)\right\rangle $
will have no physical degrees of freedom, only virtual ones that enforce
the Gauss law and connect the links.

\subsubsection{Example}

As an example~\cite{zohar_building_2016}, let us construct a very simple PEPS
as follows. At each vertex, we use a physical Hilbert space spanned
by states of the form $\left\{ \left|p\right\rangle \right\} _{p=-J}^{J}$
(eigenstates of $Q$, labeled by its eigenvalues) and virtual ones
with $\left\{ \left|v\right\rangle \right\} _{v=-j}^{j}$ (similarly,
eigenstates of the virtual electric fields $E_{0}$) where the integers
$j,J$ can be either finite or infinite. The symmetry is obtained
for
\begin{equation}
  \ket{A}=A_{ruld}^{p}\ket{p}\ket{r,u,l,d},
\end{equation}
where $\ket{r,u,l,d}$ is a product of four virtual
basis states, $\left\{ \ket{v}\right\}_{v=-j}^{j}$
, if
\begin{equation}
A_{ruld}^{p}\propto\delta\left(r+u,l+d+p\right)
\end{equation}
 -- the Kronecker delta enforces the Gauss law (in general, the tensor
elements $A_{ruld}^{p}$ should be proportional to the right combination
of Clebsch-Gordan coefficients that correspond to the correct combination
of representations~\cite{zohar_building_2016}; in this simple $U\left(1\right)$ case, the Clebsch-Gordan
coefficients are just Kronecker deltas). The gauging transformation
will simply be
\begin{equation}
\mathcal{U}_{G}\left(\ell\right)=\begin{cases}
\int d\phi e^{i\phi\left(\vb{x},1\right)E_{0}\left(r,\vb{x}\right)}\left|\phi\right\rangle \left\langle \phi\right|_{\ell} & \ell\:\mathrm{horizontal}\\
\int d\phi e^{i\phi\left(\vb{x},2\right)E_{0}\left(u,\vb{x}\right)}\left|\phi\right\rangle \left\langle \phi\right|_{\ell} & \ell\:\mathrm{vertical.}
\end{cases}
\end{equation}
Note that as a result of the gauging that identifies the physical
electric fields with the virtual ones, $\left|E\right|\leq j$, and
hence \emph{if we have a finite truncation for the virtual electric field,
the physical electric field will be equally truncated as well}.

Finally, the maximally entangled pair states on which we project will
be
\begin{equation}
\left|\omega_{\ell}\right\rangle =\begin{cases}
\underset{\left|v\right|\leq j}{\sum}\left|v\right\rangle _{\vb{x},r}\left|v\right\rangle _{\vb{x}+\vu{e}_{1},l} & \ell\:\mathrm{horizontal}\\
\underset{\left|v\right|\leq j}{\sum}\left|v\right\rangle _{\vb{x},u}\left|v\right\rangle _{\vb{x}+\vu{e}_{2},d} & \ell\:\mathrm{vertical.}
\end{cases}
\end{equation}

\section{Construction of Gauge Invariant Fermionic PEPS (fPEPS)}

While the above procedure for gauging a PEPS works for any PEPS with
a global symmetry as $\left|\psi_{0}\right\rangle $, we shall focus
now on a more particular case, of gauged Gaussian fermionic
PEPS. We begin with a fermionic free matter state with global symmetry, in accordance with the usual matter formulations in
gauge theories -- and thus we will construct a fermionic PEPS (fPEPS)
(for non fermionic lattice gauge PEPS, see~\cite{haegeman_gauging_2015} --
with matter, and~\cite{tagliacozzo_tensor_2014} -- without).

Within the class of fermionic PEPS, the states we would like to construct
and then gauge will be Gaussian fermionic PEPS~\cite{kraus_fermionic_2010}.
Gaussian states are ground states of quadratic, free, non-interacting
Hamiltonians, and thus gauging them is analogous to gauging a free
matter theory through the conventional Hamiltonian (or Lagrangian)
procedure. They are fully described in terms of their second moments
(covariance matrix) -- expectation values of fermionic bilinear, quadratic
operators; all other correlation functions are obtainable using
the Wick theorem~\cite{bravyi_lagrangian_2005}. Therefore, they make sense as
a starting point from both physical and computational grounds.

Once again, we will focus on the $U\left(1\right)$ case in $2+1$
dimensions, as it captures all the important qualitative features
without mathematical complications that have to do with choosing some
more complicated gauge group or higher dimensions. This derivation
is discussed in detail in~\cite{zohar_fermionic_2015}. The $SU(2)$ generalization
may be found in~\cite{zohar_projected_2016}, while a discussion for general
gauge groups and dimensions is in~\cite{zohar_combining_2018}.

We will hence begin with the construction of globally invariant fPEPS,
and then turn to gauging them.

\subsection{The Physical Setting}

The physical ingredients we would like to include have already been
discussed: a single fermionic mode, created by $\psi^{\dagger}\left(\vb{x}\right)$,
is defined at each vertex. We will use a staggered fermionic formulation
as before, but a slightly different one, which is obtained from the
previous one by a particle-hole transformation of the odd sublattice.
Previously, the absence of a fermion on an odd vertex represented
the presence of an anti-particle. Here, the presence of a fermion
will represent the presence of an anti-particle. This will allow us
to define things in a more translationally invariant way on the level
of states. The local fermionic charges are now
\begin{equation}
Q\left(\vb{x}\right)=\left(-1\right)^{\vb{x}}\psi^{\dagger}\left(\vb{x}\right)\psi\left(\vb{x}\right)
\end{equation}
and the Hamiltonian $H_{f}$ discussed above will transform (up to
a constant) to the superconducting form,
\begin{equation}
H_{f}=M\underset{\vb{x}}{\sum}\psi^{\dagger}\left(\vb{x}\right)\psi\left(\vb{x}\right)+\epsilon\underset{\vb{x},i=1,2}{\sum}\left(-1\right)^{\vb{x}}\left(\psi^{\dagger}\left(\vb{x}\right)\psi^{\dagger}\left(\vb{x}+\vu{e}_{i}\right)+h.c.\right).
\end{equation}

We wish to construct a state $\ket{\psi_{0}}$ with
a global $U\left(1\right)$ invariance; that is, $\ket{\psi_{0}}$
should be annihilated by
\begin{equation}
  \mathcal{Q}=\underset{\vb{x}}{\sum}Q\left(\vb{x}\right)=\underset{\vb{x}}{\sum}\left(-1\right)^{\vb{x}}\psi\dgr\left(\vb{x}\right)\psi\left(\vb{x}\right)
\end{equation}
or invariant under transformations generated by it,
\begin{equation}
  e^{i\varLambda\mathcal{Q}}\ket{\psi_{0}}=\ket{\psi_{0}}.
\end{equation}

One can also demand further symmetries, such as translation and (lattice)
rotation invariance, which we will discuss later on.

\subsection{The Local Ingredients}

Naively, we would first like to construct the states $\left|A\left(\vb{x}\right)\right\rangle $,
and their product $\underset{\vb{x}}{\otimes}\left|A\left(\vb{x}\right)\right\rangle $.
However, a product state of fermions is not well-defined, since a
tensor product factorization of a fermionic Fock space is not well
defined. One could define some ordering but we would like to avoid
that. Instead, at each vertex we will deal with an operator $A\left(\vb{x}\right)$,
involving both the physical and virtual degrees of freedom, and construct
the product state as
\begin{equation}
  \ket{A}=\underset{\vb{x}}{\prod}A\left(\vb{x}\right)\ket{\varOmega}
\end{equation}
where $\ket{\varOmega}$ is some initial state of the
system -- including the Fock vacuum of the physical fermions.

A product of fermionic operators is still not well defined without
specifying some lattice ordering, unless all the $A\left(\vb{x}\right)$
operators commute. As each of them involves degrees of freedom defined
on a different vertex, they will commute if and only if each $A\left(\vb{x}\right)$
has an even fermionic parity, which we shall thus demand. In order
to do that, the virtual degrees of freedom must be represented by
fermions as well, and we simply interpret $\left|\varOmega\right\rangle $
as the total Fock vacuum, of both physical and virtual modes (otherwise,
we must have an even local \emph{physical} parity at each vertex, which
does not make any physical sense; moreover, in our current example,
with only one physical mode per vertex, even parity with non-fermionic
virtual modes implies that the physical fermions cannot be excited).

Now we can focus on a single vertex. 
In the example we construct in the following, besides the physical mode $\psi^{\dagger}$ we introduce eight virtual (auxiliary) fermionic modes, created by
$r_{\pm}^{\dagger},u_{\pm}^{\dagger},l_{\pm}^{\dagger},d_{\pm}^{\dagger}$,
associated with the four legs attached to the vertex as before (right,
up, left, down). Out of these, we construct the virtual electric fields,
\begin{equation}
  E_{0}\left(\vb{x},z\right)=\left(-1\right)^{\vb{x}}\left(z_{+}^{\dagger}\left(\vb{x}\right)z_{+}\left(\vb{x}\right)-z_{-}^{\dagger}\left(\vb{x}\right)z_{-}\left(\vb{x}\right)\right)
\end{equation}
where $z=r,u,l,d$, and the \emph{virtual Gauss law operator}
\begin{equation}
\mathcal{G}_{0}\equiv E_{0}\left(r\right)+E_{0}\left(u\right)-E_{0}\left(l\right)+E_{0}\left(d\right)-\psi^{\dagger}\psi
\end{equation}
-- note that the staggering was incorporated into the definition of
$E_{0}\left(z,\vb{x}\right)$.

The most general Gaussian state $\left|A\right\rangle $ will be created
from the Fock vacuum $\ket{\varOmega}$ using operators
of the form
\begin{equation}
A\left(\vb{x}\right)=\exp\left(\hat{T}_{ij}\left(\vb{x}\right)\alpha_{i}^{\dagger}\left(\vb{x}\right)\alpha_{j}^{\dagger}\left(\vb{x}\right)\right)
\end{equation}
where $\left\{ \alpha_{i}^{\dagger}\right\} $ include
all the possible creation operators, either physical or virtual.

We wish to construct $A\left(\vb{x}\right)$ which are invariant
under this virtual Gauss law operator,
\begin{equation}
e^{i\varLambda\left(\vb{x}\right)\mathcal{G}_{0}\left(\vb{x}\right)}A\left(\vb{x}\right)e^{-i\varLambda\left(\vb{x}\right)\mathcal{G}_{0}\left(\vb{x}\right)}=A\left(\vb{x}\right)
\end{equation}
In order to do that, let us sort our modes to positive and negative
ones, depending on the sign of the phase which is put on them by this
transformation. The negative ones, $\left\{ a_{i}^{\dagger}\right\} $
are the ones whose number operator appears in $\mathcal{G}_{0}$ with
a minus sign on an even vertex, and therefore
\begin{equation}
  e^{i\varLambda\mathcal{G}_{0}}a_{i}\dgr e^{-i\varLambda\mathcal{G}_{0}}=e^{-i\varLambda}a_{i}^{\dagger},
\end{equation}
while the positive ones, $\left\{ b_{i}^{\dagger}\right\} $ are represented
by number operators with a plus sign in front, giving rise to
\begin{equation}
e^{i\varLambda\mathcal{G}_{0}}b_{i}\dgr e^{-i\varLambda\mathcal{G}_{0}}=e^{i\varLambda}b_{i}\dgr.
\end{equation}
On an even vertex, the negative modes are $\left\{ a_{i}^{\dagger}\right\} =\left\{ \psi^{\dagger},r_{-}^{\dagger},u_{-}^{\dagger},l_{+}^{\dagger},d_{+}^{\dagger}\right\} $
and the positive ones - $\left\{ b_{i}^{\dagger}\right\} $$=\left\{ r_{+}^{\dagger},u_{+}^{\dagger},l_{-}^{\dagger},d_{-}^{\dagger}\right\} $.
On the odd sublattice, $\left\{ a_{i}^{\dagger}\right\} =\left\{ \psi^{\dagger},r_{+}^{\dagger},u_{+}^{\dagger},l_{-}^{\dagger},d_{-}^{\dagger}\right\} $
and the positive ones - $\left\{ b_{i}^{\dagger}\right\} $$=\left\{ r_{-}^{\dagger},u_{-}^{\dagger},l_{+}^{\dagger},d_{+}^{\dagger}\right\} $.

It is easy to convince ourselves that the symmetry condition is satisfied
if and only if the constructed state has the form
\begin{equation}
A\left(\vb{x}\right)=\exp\left(T_{ij}\left(\vb{x}\right)a_{i}^{\dagger}\left(\vb{x}\right)b_{j}^{\dagger}\left(\vb{x}\right)\right),
\end{equation}
where $T_{ij}$ is a $5\times4$ matrix; the first row corresponds
to the coupling virtual and physical modes, and the remaining square
block, $\tau$ couples the virtual modes among themselves. A minimal
choice is $T_{ij}\left(\vb{x}\right)=T_{ij}$. Out of it we can
construct a translationally invariant fPEPS (that will become charge
conjugation invariant once we gauge): the staggering implies that
when one translates by one lattice site, $a_{i}^{\dagger}\longleftrightarrow b_{i}^{\dagger}$
for the virtual modes (translation by one lattice site for staggered
fermions is charge conjugation), and thus the square, virtual block
$\tau$ has to be antisymmetric.

We obtain, for every $\vb{x}$, that
\begin{equation}
e^{i\varLambda\mathcal{G}_{0}}Ae^{-i\varLambda\mathcal{G}_{0}}=W_{r}W_{u}W_{l}^{\dagger}W_{d}^{\dagger}V^{\dagger}AVW_{d}W_{l}W_{u}^{\dagger}W_{r}^{\dagger}
\end{equation}
-- the PEPS condition for global symmetry, as in the non-fermionic
case.
\begin{center}
  \includegraphics{globalsym.pdf}
\end{center}

\subsection{Contracting the Globally Invariant fPEPS}

Once again, to avoid the product of fermionic states, we use operators
instead of states for the contraction. Since we wish our final state
to be Gaussian, we will use Gaussian projectors for that. On a horizontal
link, we define the (unnormalized) projector
\begin{equation}
  \omega_{\vb{x},1}=\exp\left(l_{+}^{\dagger}\left(\vb{x}+\vu{e}_{1}\right)r_{-}^{\dagger}\left(\vb{x}\right)+l_{-}^{\dagger}\left(\vb{x}+\vu{e}_{1}\right)r_{+}^{\dagger}\left(\vb{x}\right)\right)\Omega_{\ell}\exp\left(r_{-}\left(\vb{x}\right)l_{+}\left(\vb{x}+\vu{e}_{1}\right)+r_{+}\left(\vb{x}\right)l_{-}\left(\vb{x}+\vu{e}_{1}\right)\right)
\end{equation}
and on a vertical one,
\begin{equation}
  \omega_{\vb{x},2}=\exp\left(u_{+}^{\dagger}\left(\vb{x}\right)d_{-}^{\dagger}\left(\vb{x}+\vu{e}_{2}\right)+u_{-}^{\dagger}\left(\vb{x}\right)d_{+}^{\dagger}\left(\vb{x}+\vu{e}_{2}\right)\right)\Omega_{\ell}\exp\left(d_{-}\left(\vb{x}+\vu{e}_{2}\right)u_{+}\left(\vb{x}\right)+d_{+}\left(\vb{x}+\vu{e}_{2}\right)u_{-}\left(\vb{x}\right)\right),
\end{equation}
where in both cases $\Omega_{\ell}$ is the projector to the vacuum
states of the modes on the link's edges.

It is easy to see that for horizontal links
\begin{equation}
W_{l}W_{r}\omega_{\ell}=\omega_{\ell}=\omega_{\ell}W_{r}W_{l}
\end{equation}
while for vertical ones
\begin{equation}
W_{u}W_{d}\omega_{\ell}=\omega_{\ell}=\omega_{\ell}W_{d}W_{u}
\end{equation}
-- this can be recognized as the usual projectors symmetry conditions
for a PEPS, but different than the ones previously used, due to the
staggering.
\begin{center}
  \includegraphics{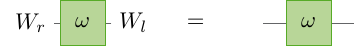}
\end{center}

The fPEPS is then given by
\begin{equation}
\left|\psi_{0}\right\rangle =\underset{\text{Step 4}}{\underbrace{\left\langle \varOmega_{v}\right|\underset{\ell}{\prod}\omega_{\ell}}}
\underset{\text{Step 1}}{\underbrace{\underset{\vb{x}}{\prod}A\left(\vb{x}\right)\left|\varOmega\right\rangle}},
\end{equation}
where $\left|\varOmega_{v}\right\rangle $ is the virtual vacuum.
The global transformations take the form $e^{i\varLambda\mathcal{Q}}=\underset{\vb{x}\:\mathrm{even}}{\prod}V\left(\vb{x}\right)\underset{\vb{x}\:\mathrm{odd}}{\prod}V^{\dagger}\left(\vb{x}\right)$
-- but as this staggering has already been taken care of, $e^{i\varLambda\mathcal{Q}}\left|\psi_{0}\right\rangle =\left|\psi_{0}\right\rangle $.

To avoid confusing notations due to the staggering, we show this
only graphically:
\begin{center}
  \scalebox{0.5}{
    \includegraphics{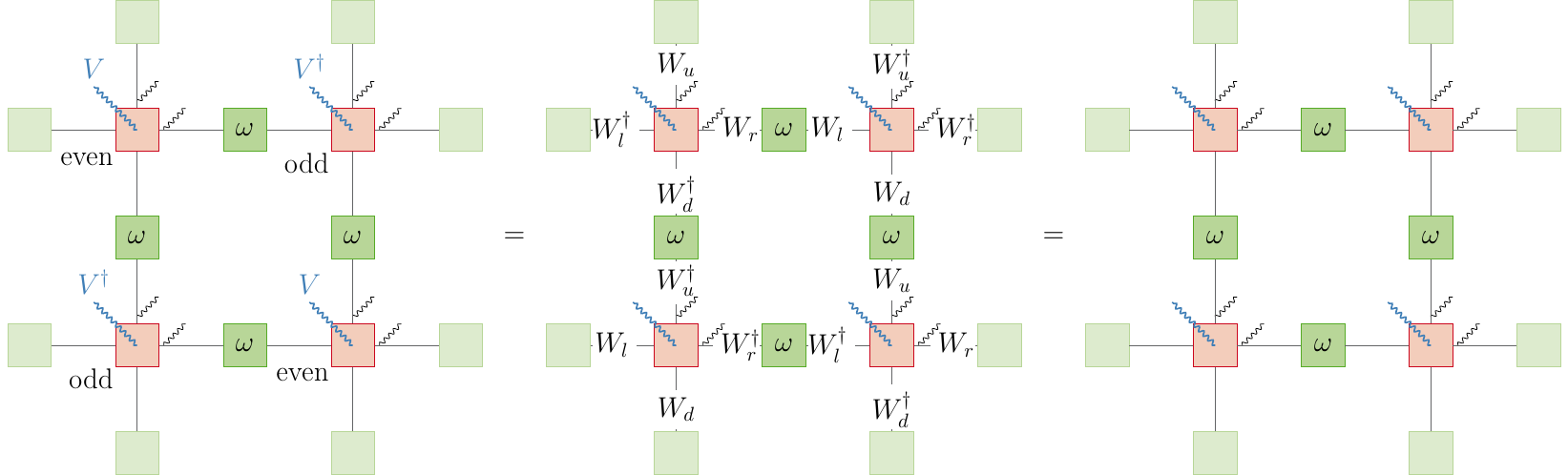}
  }
\end{center}

\subsection{Gauging the PEPS}

We go on to gauging the state $\left|\psi\right\rangle $. We introduce
gauge field Hilbert spaces on the links, but how shall we relate them
to the fermions? We simply have to modify the gauged PEPS we constructed
previously to the fermionic case:
\begin{equation}
\left|\psi\right\rangle =
\underset{\text{Step 4}}{\underbrace{\left\langle \varOmega_{v}\right|\underset{\ell}{\prod}\omega_{\ell}}}
\underset{\text{Step 3}}{\underbrace{\underset{\ell}{\prod}\mathcal{U}_{G}\left(\ell\right)}}
\underset{\vb{x}}{\prod}\underset{\text{Step 2}}{\underbrace{\underset{\text{Step 1}}{\underbrace{A\left(\vb{x}\right)}}\left|0\right\rangle _{\vb{x}}}}\left|\varOmega\right\rangle,
\end{equation}
where $\mathcal{U}_{G}\left(\ell\right)$ is a gauging transformation
that entangles the virtual fermion at the beginning of the link $\ell$
with the physical gauge field we introduce on it, i.e. it rotates
the virtual fermion with respect to the gauge field. It is defined
as
\begin{equation}
\mathcal{U}_{G}\left(\ell\right)=\begin{cases}
\int d\phi e^{i\left(-1\right)^{\vb{x}}\phi\left(\vb{x},1\right)E_{0}\left(r,\vb{x}\right)}\left|\phi\right\rangle \left\langle \phi\right|_{\ell} & \ell\:\mathrm{horizontal}\\
\int d\phi e^{i\left(-1\right)^{\vb{x}}\phi\left(\vb{x},2\right)E_{0}\left(u,\vb{x}\right)}\left|\phi\right\rangle \left\langle \phi\right|_{\ell} & \ell\:\mathrm{vertical}.
\end{cases}
\end{equation}

We define
\begin{equation}
A_{G}\left(\vb{x}\right)=\mathcal{U}_{G}\left(\vb{x},1\right)\mathcal{U}_{G}\left(\vb{x},2\right)A\left(\vb{x}\right)\left|0\right\rangle _{\vb{x}}.
\end{equation}
What are the symmetry conditions satisfied by this object? It is easy
to see that for an even vertex $\vb{x}$,
\begin{equation}
V_{r,u}A_{G}\left(\vb{x}\right)\left|\Omega\right\rangle =W_{r,u}A_{G}\left(\vb{x}\right)\left|\Omega\right\rangle
\end{equation}
\begin{center}
  \includegraphics{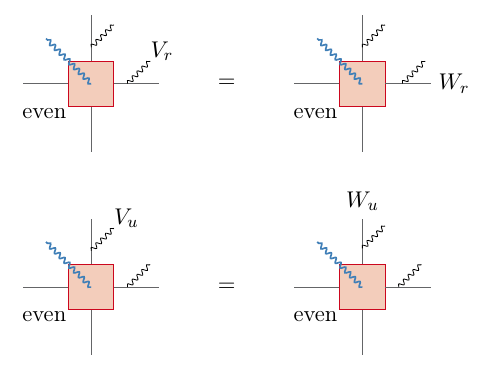}
\end{center}
while for an odd one
\begin{equation}
V_{r,u}A_{G}\left(\vb{x}\right)\left|\Omega\right\rangle =W_{r,u}^{\dagger}A_{G}\left(\vb{x}\right)\left|\Omega\right\rangle
\end{equation}
\begin{center}
  \includegraphics{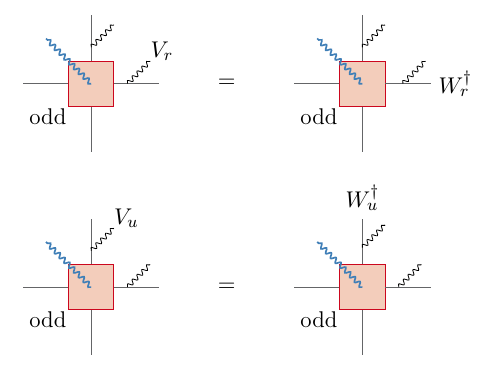}
\end{center}
-- the properties that allow us to gauge, just like in~\eqref{G2}. From
these two equations, and the symmetry condition of $A\left(\vb{x}\right)$,
we obtain that for an even vertex,\\
\begin{minipage}{\textwidth}
  \begin{equation}
    V_{r}V_{u}V^{\dagger}A_{G}\left(\vb{x}\right)\left|\Omega\right\rangle =W_{l}W_{d}A_{G}\left(\vb{x}\right)\left|\Omega\right\rangle
  \end{equation}
  \begin{center}
    \includegraphics{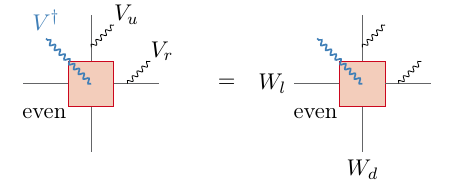}
  \end{center}
\end{minipage}
while for an odd one,\\
\begin{minipage}{\textwidth}
  \begin{equation}
    V_{r}V_{u}VA_{G}\left(\vb{x}\right)\left|\Omega\right\rangle =W_{l}^{\dagger}W_{d}^{\dagger}A_{G}\left(\vb{x}\right)\left|\Omega\right\rangle
  \end{equation}
  \begin{center}
    \includegraphics{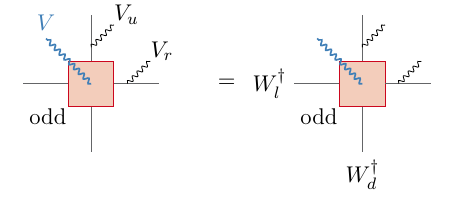}
  \end{center}
\end{minipage}

This allows us to verify that $\left|\psi\right\rangle $ is gauge
invariant, i.e. that for an arbitrary \textbf{$\vb{x}_{0}$},
\begin{equation}
  \begin{aligned}
    e^{i\varLambda\mathcal{G}\left(\vb{x}_{0}\right)}\left|\psi\right\rangle&=V_{r}\left(\vb{x}_{0}\right)V_{u}\left(\vb{x}_{0}\right)V_{l}^{\dagger}\left(\vb{x}_{0}-\vu{e}_{1}\right)V_{d}^{\dagger}\left(\vb{x}_{0}-\vu{e}_{2}\right)V^{\dagger}\left(\vb{x}_{0}\right)\ket{\psi} \\
  &=\left|\psi\right\rangle
  \end{aligned}
\end{equation}
Once again we show it pictorially, for the two possible parities of
the vertex around which the state is transformed.
First, even,
\begin{center}
  \includegraphics[scale=0.5]{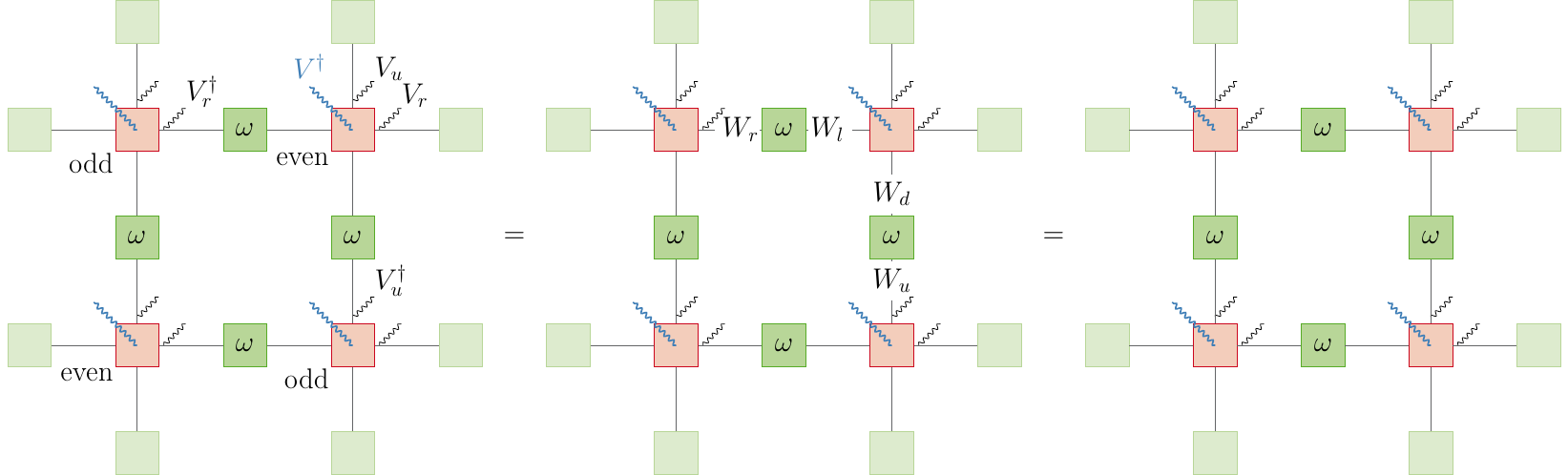}
\end{center}
and finally odd,
\begin{center}
  \includegraphics[scale=0.5]{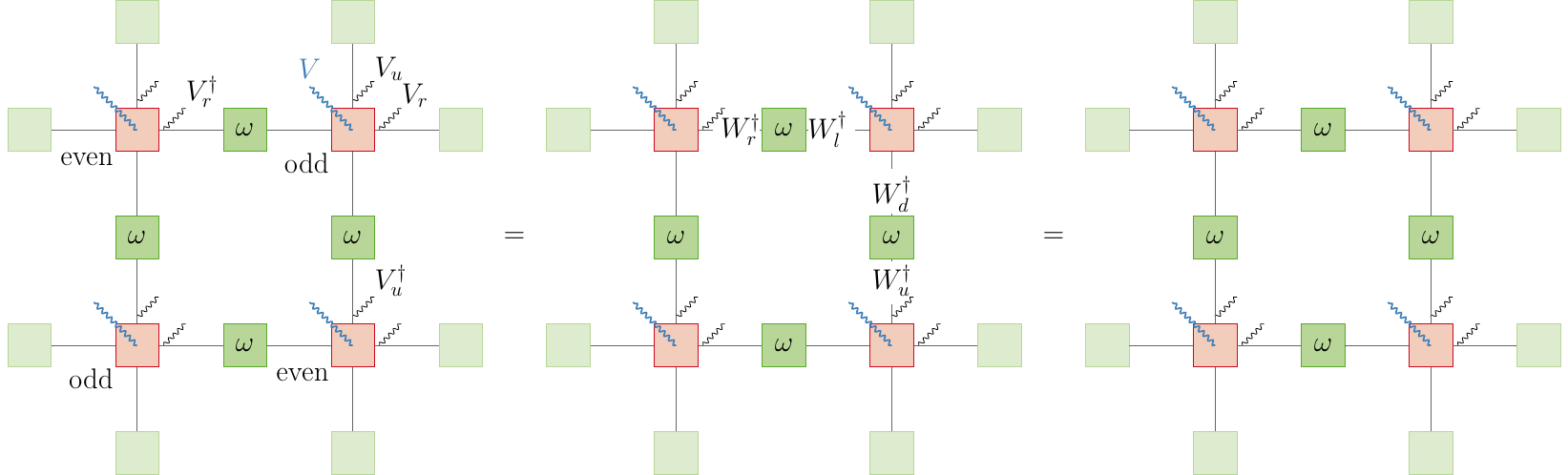}
\end{center}

Another thing that has to be mentioned, is that in this method the
physical electric field is truncated, and cannot take any integer
eigenvalue, but rather only, in this case, $0,\pm1$. This is once
again due to the fact that in the gauging process (the action of $\mathcal{U}_{G}$)
transfers the state of the virtual electric field to the physical
one, and the virtual electric fields, constructed out of fermions
in this particular manner (as the difference between two fermionic
number operators) only take these values. In order to increase the
truncation, one has to add more virtual fermions on each leg, such
that higher eigenvalues of $E_{0}$ are obtained. This is called,
in usual PEPS terms, increasing the \emph{bond dimension}, and it also allows
one, in general, to increase the number of parameters the PEPS depends
on.

\subsection{Other Symmetries}

So far, we have constructed a fermionic Gaussian PEPS with a global
$U\left(1\right)$ symmetry, and made it local using our gauging procedure.
What can we say about other symmetries obeyed by the state?

We have already commented how to encode translational invariance into
the globally invariant state $\left|\psi_{0}\right\rangle $, and
that a translation by one lattice site, which is, for staggered fermions,
charge conjugation, requires exchanging $a_{i}^{\dagger}\longleftrightarrow b_{i}^{\dagger}$.
If we consider the staggering in the definition of $\mathcal{U}_{G}\left(\ell\right)$,
we see that the transformation of the gauge field that leaves $\left|\psi\right\rangle $
invariant is translation + inversion of sign,
\begin{equation}
\phi\left(\vb{x},i\right)\longrightarrow-\phi\left(\vb{x}+\vu{e}_{j},i\right)
\end{equation}
-- which is exactly charge conjugation! Of course, it makes no sense
to talk about simple translational invariance when we discuss staggered
fermions, where the unit cell includes more than one vertex. If we
consider translation by two sites, on the other hand, we have simple
translational invariance,
\begin{equation}
\phi\left(\vb{x},i\right)\longrightarrow\phi\left(\vb{x}+2\vu{e}_{j},i\right)
\end{equation}
as expected for such a staggered system.

Another symmetry that could be considered is rotation. First of all,
for the globally invariant state without the gauge field. For that,
one has to define the right transformation properties of all fermions
under the rotation. This involves not only changing the coordinate
respectively, but also requires more. As we discuss fermions, a complete
rotation of the physical fermion must put a sign on it. Therefore,
the basic $\pi/2$ rotation, realized by an operator $\mathcal{U}_{R}$,
must give rise to
\begin{equation}
\mathcal{U}_{R}\psi^{\dagger}\left(\vb{x}\right)\mathcal{U}_{R}^{\dagger}=\eta_{p}\psi^{\dagger}\left(R\vb{x}\right)
\end{equation}
with $\eta_{p}^{4}=-1$; $R\vb{x}$ is the rotated coordinate.

The virtual modes around each vertex must be exchanged by such a transformation,
up to possible phases, keeping positive modes positive and negative
modes negative. That is, we need to have, for the virtual components,
\begin{equation}
  \begin{aligned}
    \mathcal{U}_{R}a_{i}\dgr\left(\vb{x}\right)\mathcal{U}_{R}\dgr&=M_{ij}^{A}a_{j}\dgr\left(R\vb{x}\right)\\
    \mathcal{U}_{R}b_{i}\dgr\left(\vb{x}\right)\mathcal{U}_{R}\dgr&=M_{ij}^{B}b_{j}\dgr\left(R\vb{x}\right)
  \end{aligned}
\end{equation}
with rotation matrices that contain only one nonzero element per row
and column, and that has to be a phase. The phase must be chosen such
that the product of projectors $\underset{\ell}{\prod}\omega_{\ell}$
is invariant.

After gauging, one simply has to add the rotation rules for the gauge field. 
Once these two requirements (translation and
rotation invariance) are taken care of, and we remove redundancies
having to do with phase symmetries of the virtual fermions in
which the physical fermions do not participate.
We arrive at a final characterization~\cite{zohar_fermionic_2015},
\begin{equation}
  T=\left(\begin{array}{cccc}
  t & \eta_{p}^{2}t & \eta_{p}t & \eta_{p}^{3}t\\
  0 & y & z/\sqrt{2} & z/\sqrt{2}\\
  -y & 0 & -z/\sqrt{2} & z/\sqrt{2}\\
  -z/\sqrt{2} & z/\sqrt{2} & 0 & y\\
  -z/\sqrt{2} & -z/\sqrt{2} & -y & 0
  \end{array}\right),
\end{equation}
where $t\geq0$ and $y,z\in\mathbb{C}$. $t=0$ is of course meaningless
in the global case, but in the local case it gives rise to pure gauge
states.

\subsection{Generalizations}

One can generalize the gauging prescription for PEPS, and in particular
Gaussian fPEPS, described above, in the following ways:
\begin{enumerate}
\item \emph{Different spatial dimensions}: one has to introduce more similar
virtual legs, and apply the gauging transformation only in the outgoing
directions. A similar procedure holds for other geometries too.
\item \emph{Different gauge groups}: similar procedure applies to further gauge
groups (see~\cite{zohar_projected_2016} for an explicit $SU\left(2\right)$
construction, and~\cite{zohar_combining_2018} for a general formulation). In
general, for non-Abelian groups, the symmetry conditions become more
complicated technically, as there is difference between left and right
group transformations~\cite{kogut_hamiltonian_1975,zohar_formulation_2015} (the electric field on the link differs from
one edge to another -- it has both \emph{left} and \emph{right} fields,
whose difference manifests the fact that a non-Abelian gauge fields,
unlike Abelian ones, carry charge -- e.g. gluons vs. photons).
\item \emph{Larger bond dimensions}: as explained before, one may add more virtual
fermions in order to enlarge the dimension of the virtual Hilbert
spaces and introduce more parameters on which the state depends. This,
however, must be done carefully in a way that respects the symmetry
(only complete irreducible multiplets can be included -- and also multiple
copies thereof). This increases the truncation of the virtual Hilbert
spaces, and hence of the physical Hilbert space on the links as well
(electric field truncation).
\end{enumerate}

\subsection{Analytical (Toy Model) Results}
Which type of physics is manifested by "toy" states, with
the minimal bond dimension (truncated electric fields), as described above?

One way to study that analytically is to put the system on a cylinder
-- with one compact and one open dimension -- and define an object called
the PEPS transfer matrix, similar to transfer matrix in statistical
mechanics~\cite{yang_chiral_2015}. The transfer matrix "transfers" information
from layer to layer of the cylindrical system, along the open direction. If the spectrum of the transfer
matrix is gapped, correlation functions with respect to the PEPS show
an exponential decay.
%The correlation function can be calculated as
%\begin{equation}
%  \expval{O(y)O(y+L)}\sim \overbrace{T_{aa'}T_{a'a''}\dots T_{b'b}}^{L-1\,\text{copies}\sim L}.
%\end{equation}
%The construction of the transfer matrix is depicted in Figure~\ref{fig:peps_torus}.
%\begin{figure}[h]
%  \centering
%  \includegraphics[scale=0.45]{cyl.pdf}
%  %FIXME: Enter caption
%  \caption{Enter Caption}
%  \label{fig:peps_torus}
%\end{figure}

A conventional way to study such PEPS is therefore to compute the
lowest eigenvalues of the transfer matrix for some different choices
of the PEPS parameters, and see whether it is gapped or not -- implying
that the PEPS could be the ground state of a gapped Hamiltonian, or
not. This allows one to draw a ``phase diagram of states'' in this
parameter space -- that is, a phase diagram for the set of Hamiltonians
of which the set of PEPS obtained by varying the parameters, are ground
states (usually referred to as parent Hamiltonians). Parameter space
surfaces on which the transfer matrix is gapless are suspected as
phase boundaries.

%In a case of a gauge invariant PEPS, such as ours, a two-point correlation
%function for two operators in two different rows, can be nonzero only
%if both operators are separately gauge invariant. Otherwise, a gauge
%field string must connect them. Therefore another type of a transfer
%matrix, with a $U$ operator on it, is also physically meaningful
%-- for example, for the computation of the expectation value of mesonic
%operators:
%\begin{equation}
%  \expval{\Psi\dgr(y)\left( \prod_{y\leq z<y+L}U(z) \right)\Psi(y+L)}.
%\end{equation}
%This case is represented in Figure~\ref{fig:peps_torus_op}.
%
%\begin{figure}[h]
%  \centering
%  \includegraphics[scale=0.45]{cylst.pdf}
%  %FIXME: Enter caption
%  \caption{Add caption}
%  \label{fig:peps_torus_op}
%\end{figure}

One could then draw a phase diagram in this method, and calculate
the expectation values of some significant physical operators -- order
parameters -- within the phases (gapped regions). Such studies, with
Wilson loops~\cite{wilson_confinement_1974}, for $U\left(1\right)$ states~\cite{zohar_fermionic_2015},
have shown both confining and non-confining gapped phases for static
charges, in the pure gauge case. Similar $SU\left(2\right)$ studies
\cite{zohar_projected_2016} have shown a gapped deconfining phase as well as
a gapless phase with perimeter-law decay of the area law -- a hint
of a possible Higgs phase~\cite{fradkin_phase_1979}.

Another way to detect possible phase transitions is to look for virtual
symmetries: symmetries in which the physical degrees of freedom do
not participate. In the $SU\left(2\right)$ case~\cite{zohar_projected_2016},
for example, such a particle-hole symmetry exists for the virtual
fermions. This symmetry allows one to construct a parameter space
transformation that connects states which are physically identical
but constructed using different parameters. The fixed lines of this
transformation are reasonably suspected as phase boundaries -- and
indeed, when those were compared to transfer matrix calculations,
they gave rise to the same results.

\subsection{Contracting Gauged Gaussian Fermionic PEPS: Combining Monte-Carlo
with Tensor Network Methods}

In many cases, tensor network states are used as ansatz states in
variational calculations. While this has been done quite successfully
for one dimensional systems, with DMRG (density matrix renormalization
group) techniques~\cite{white_density_1992}, the higher dimensional generalizations
are problematic as they do not scale well~\cite{verstraete_matrix_2008}.

From the lattice gauge theory point of view, the widely used Monte-Carlo
methods suffer from the sign problem~\cite{troyer_computational_2005} in several
physically interesting scenarios (e.g. finite chemical potential).
They also do not allow one to consider directly unitary time evolution,
as they are formulated in a Euclidean spacetime.

Indeed, in one space dimension, tensor network numerical studies have shown
remarkable results, even in otherwise problematic scenarios
\cite{byrnes_density_2002,
  sugihara_matrix_2005,
  banuls_mass_2013,
  buyens_matrix_2014,
  silvi_lattice_2014,
  rico_tensor_2014,
  saito_temperature_2014,
  kuhn_non-abelian_2015,
  banuls_thermal_2015,
  banuls_chiral_2016,
  pichler_real-time_2016,
  silvi_finite-density_2017,
  milsted_matrix_2016,
  buyens_hamiltonian_2016,
  banuls_multi-flavor_2016,
  buyens_real-time_2017,
  banuls_density_2017,
  banuls_mari_carmen_towards_2017,
  banuls_efficient_2017,
  carmen_banuls_tensor_2018,
  silvi_tensor_2019,
  banuls_tensor_2018,
  silvi_tensor_2019-1,
  tschirsich_phase_2019}.
However, what about higher dimensional systems?

Gaussian fermionic PEPS suggest a way to circumvent these problems.
As we shall see, one may calculate, using Monte-Carlo methods, the
expectation values of physical observables with respect to gauged
Gaussian PEPS, exploiting the simple and efficient Gaussian formalism,
based on Gaussian integration and matrix operations~\cite{bravyi_lagrangian_2005}.
Such calculations do not depend on the dimension of the system, overcoming
the first problem. They also do not encounter the sign problem, as
they are carried out in the Hamiltonian, Hilbert space formulation (expectation
values of physical operators) and not in the Wick rotated path integral
formulation that encounters it.

For a general discussion see~\cite{zohar_combining_2018}; here, we shall briefly
review the simple $U\left(1\right)$ case.

We have already mentioned that the Hilbert space of a fermionic lattice
gauge theory $\mathcal{H}$, with no static charges, could be embedded
in the product space
\begin{equation}
  \mathcal{H} \subset \mathcal{H}_{\mathrm{g}} \times \mathcal{H}_{\mathrm{f}},
\end{equation}
where $\mathcal{H}_{\mathrm{g}}$ is the gauge field Hilbert space,
covering all links, and $\mathcal{H}_{\mathrm{f}}$ is the fermionic
Fock space, covering all vertices. The most general state in this
Hilbert space may be expanded as
\begin{equation}
  \ket{\psi}=\int\mathcal{D}\Phi\ket{\Phi}\ket{\psi\left(\Phi\right)},
\end{equation}
where $\ket{\Phi}=\underset{\ell}{\otimes}\ket{\phi_{\ell}}\in\mathcal{H}_{\mathrm{g}}$
is some gauge field configuration state on all the links, and $\mathcal{D}\Phi=\underset{\ell}{\prod}d\phi_{\ell}$. 
$\ket{\psi\left(\Phi\right)}\in\mathcal{H}_{\mathrm{f}}$ is a general fermionic state, that depends on the gauge field configuration
as a parameter, and is in general not normalized. 
The state $\ket{\psi}$ could be very general, not necessarily a PEPS and even not manifesting any symmetry;
of course, we are interested in the case when it is gauge invariant.
Then, $\ket{\psi\left(\Phi\right)}$ is a state of fermions
coupled to an external, static $U\left(1\right)$ field with configuration
$\Phi$.

When $\ket{\psi}$ is a gauged Gaussian fermionic PEPS,
the states $\ket{\psi\left(\Phi\right)}$ are merely
fermionic Gaussian states, and therefore admit a very simple description
and analytical treatment using the Gaussian formalism~\cite{bravyi_lagrangian_2005}.
This allows us to efficiently calculate the expectation values of
physical operators for them.

Consider the (square) norm of the state. As the configuration states
$\ket{\Phi}$ form a complete set, it can be simply
expressed as
\begin{equation}
\left\langle \psi|\psi\right\rangle =\int\mathcal{D}\Phi\left\langle \psi\left(\Phi\right)|\psi\left(\Phi\right)\right\rangle \equiv Z
\end{equation}
-- an integral over squares of norms of Gaussian states, $\left\langle \psi\left(\Phi\right)|\psi\left(\Phi\right)\right\rangle $,
whose computation is simple and numerically efficient. 
They are all positive definite, and thus the function
\begin{equation}
p\left(\Phi\right)=\left\langle \psi\left(\Phi\right)|\psi\left(\Phi\right)\right\rangle /Z
\end{equation}
is a probability density function in the space of gauge field configurations,
and $Z$ is the partition function associated with it.

Now we are ready to compute expectation values. Consider first the
Wilson loop~\cite{wilson_confinement_1974}, an oriented product of group element operators
along some closed path $\mathcal{C}$:
\begin{equation}
W\left(\mathcal{C}\right)=\underset{\ell\in\mathcal{C}}{\prod}U\left(\ell\right),
\end{equation}
where $\ell$ is now an oriented link.
We define orientation with respect to the Gauss law.
If the link is following an outgoing leg (up, right), it is considered \emph{forward}.
The other two links (down, left) are \emph{backward}.
If its orientation is backwards one has to use $U\dgr$ on that link, such
that the operator $W\left(\mathcal{C}\right)$ is gauge invariant.
The configuration states are eigenstates of $W\left(\mathcal{C}\right)$,
\begin{equation}
  W\left(\mathcal{C}\right)\left|\Phi\right\rangle =\exp\left(i\underset{\ell\in\mathcal{C}}{\sum}\phi\left(\ell\right)\right)\left|\Phi\right\rangle
\end{equation}
where now a minus sign is understood in front of the phase of a backward link. 
We therefore obtain that
\begin{equation}
  \left\langle W\left(\mathcal{C}\right)\right\rangle =\left\langle \psi\right|W\left(\mathcal{C}\right)\left|\psi\right\rangle /Z=\int\mathcal{D}\Phi p\left(\Phi\right)\exp\left(i\underset{\ell\in\mathcal{C}}{\sum}\phi\left(\ell\right)\right),
\end{equation}
-- which can be computed using Monte-Carlo.

Let us consider another type of physical operator -- a mesonic string,
an open gauge field string enclosed by fermionic operators, e.g.
\begin{equation}
M\left(\vb{x},\vb{y},\mathcal{C}\right)=\psi^{\dagger}\left(\vb{x}\right)\underset{\ell\in\mathcal{C}}{\prod}U\left(\ell\right)\psi\left(\vb{y}\right)
\end{equation}
where now $\mathcal{C}$ is an open path connecting the endpoints
$\vb{x},\vb{y}$. In this case, we obtain that
\begin{equation}
\left\langle M\left(\vb{x},\vb{y},\mathcal{C}\right)\right\rangle =\int\mathcal{D}\Phi p\left(\Phi\right)e^{i\underset{\ell\in\mathcal{C}}{\sum}\phi\left(\ell\right)}\frac{\left\langle \psi\left(\Phi\right)\right|\psi^{\dagger}\left(\vb{x}\right)\psi\left(\vb{y}\right)\left|\psi\left(\Phi\right)\right\rangle }{\left\langle \psi\left(\Phi\right)|\psi\left(\Phi\right)\right\rangle }
\end{equation}
-- this integral involves the expectation value of a quadratic fermionic
operator with respect to a fermionic Gaussian state -- $\left\langle \psi^{\dagger}\left(\vb{x}\right)\psi\left(\vb{y}\right)\right\rangle _{\Phi}=\frac{\left\langle \psi\left(\Phi\right)\right|\psi^{\dagger}\left(\vb{x}\right)\psi\left(\vb{y}\right)\left|\psi\left(\Phi\right)\right\rangle }{\left\langle \psi\left(\Phi\right)|\psi\left(\Phi\right)\right\rangle }$
-- which is very simple to calculate in the Gaussian formalism using
covariance matrix elements of $\left|\psi\left(\Phi\right)\right\rangle$. 
Therefore, $\left\langle M\left(\vb{x},\vb{y},\mathcal{C}\right)\right\rangle $
can be calculated too using Monte Carlo integration of quantities
obtained from Gaussian calculations.

Similar results are obtained for other operators -- such as ones involving
electric fields~\cite{zohar_combining_2018}. In all cases, the quantum nature
of the gauge field Hilbert space is handled before the integration
takes place, leaving traces of the gauge field only in the form of
integration variables. The quantum nature of the fermionic part is
taken care of through the Gaussian formalism, thanks to which everything
can be computed efficiently -- as functions of the gauge field configurations.
Eventually one is left with a $\int \mathcal{D}\Phi$ integral over the gauge field configurations
that can be computed with Monte-Carlo. This opens the way for using
such states, with higher bond dimensions, as varitational guesses
for ground states of lattice gauge theory Hamiltonians, for example.

\section{The Opposite Process: Eliminating the Fermions}

\subsection{Solving Gauss Law for the Matter Field}

We have argued that the Gauss law~\eqref{Gc} is problematic to solve
for the gauge field, since it is a differential equation with no unique
solution, if the spatial dimension is more than one (and in the one
dimensional case, the solution is unique but nonlocal). However, note that if
we see it as an equation for the matter, the setting is completely
different: then, it is a very simple algebraic equation, in which
the charge (or its density) -- not a vector quantity -- is already explicitly
and uniquely given as a function of the divergence of the electric
field.

Therefore, in cases where each local charge $Q\left(\vb{x}\right)$ eigenstate is obtained uniquely by a single configuration of the matter degree of freedom at $\vb{x}$, one could indeed write down an "opposite" unitary transformation;
that will take a gauge invariant state or Hamiltonian, with both matter
and gauge fields as quantum degrees of freedom, and decouple the
matter using the Gauss laws. This produces a state, or a Hamiltonian,
without the symmetry, but also without the Gauss law constraints;
with the same physical information, and effectively less degrees of
freedom. It is, in fact, a controlled operation that changes the state
of the matter at each vertex, controlled by the values of the electric
field on the links around it. It is a well known procedure for Higgs matter fields,
known as fixing the \emph{unitary gauge}.

\subsection{The Unitary Gauge of Higgs Theories}

Let us consider a lattice formulation of the Higgs mechanism, similar
to that of~\cite{fradkin_phase_1979}. On each vertex,
we have a complex scalar field $\Phi\left(\vb{x}\right)$ which
may be expanded in terms of two bosonic modes. We will write it in
a polar form,
\begin{equation}
  \Phi\left(\vb{x}\right)=R\left(\vb{x}\right)e^{i\theta\left(\vb{x}\right)}.
\end{equation}
In the usual quasi-classical treatment of the Higgs mechanism, $R\left(\vb{x}\right)$
is the Higgs field, and $\theta\left(\vb{x}\right)$ is the Goldstone
mode, which are independent degrees of freedom, satisfying
\begin{equation}
  \comm{R\left(\vb{x}\right)}{\theta\left(\vb{y}\right)}=0.
\end{equation}

The phase $\theta\left(\vb{x}\right)$ is canonically conjugate to the local charge $Q\left(\vb{x}\right)$, having a non bounded integer spectrum, raised by $e^{i \theta\left(\vb{x}\right)}$, and completely decopuled from the radial degree of freedom $R\left(\vb{x}\right)$.
In~\cite{fradkin_phase_1979}, the Higgs field is frozen -- $R\left(\vb{x}\right)=R_{0}$,
and we will do it too, as it has no relevance for us.
 On the links,
we have the usual $U\left(1\right)$ gauge field Hilbert space introduced
above (note that in this case, the link and matter Hilbert spaces
are the same).

The interaction Hamiltonian between the matter and gauge fields takes
the form
\begin{equation}
\widetilde{H}=\epsilon\underset{\vb{x},i=1,2}{\sum}\left(e^{i\theta\left(\vb{x}\right)}U\left(\vb{x},i\right)e^{-i\theta\left(\vb{x}+\vu{e}_{i}\right)}+h.c.\right)=2\epsilon\underset{\vb{x},i=1,2}{\sum}\cos\left(\theta\left(\vb{x}\right)+\phi\left(\vb{x},i\right)-\theta\left(\vb{x}+\vu{e}_{i}\right)\right).
\end{equation}

As usual, in the Higgs mechanism discussion, we fix the gauge to the
\emph{unitary gauge} \cite{fradkin_phase_1979}: perform a pure-gauge transformation (one that
only acts on the gauge field degrees of freedom), in which the Goldstone
modes are absorbed by the gauge field, making it massive. The transformation
is
\begin{equation}
\phi\left(\vb{x},i\right)\longrightarrow\phi\left(\vb{x},i\right)-\theta\left(\vb{x}\right)+\theta\left(\vb{x}+\vu{e}_{i}\right).
\end{equation}

How is it implemented quantum mechanically? It is a straightforward
exercise to convince ourselves that it takes the form
\begin{equation}
  \begin{aligned}
  \mathcal{U}_{H}&=\underset{\vb{x}}{\prod}\mathcal{U}_{H}\left(\vb{x}\right)\\
  \mathcal{U}_{H}\left(\vb{x}\right)&=e^{iG\left(\vb{x}\right)\theta\left(\vb{x}\right)}\\
  &=e^{i\left(E\left(\vb{x},1\right)+E\left(\vb{x},2\right)-E\left(\vb{x}-\vu{e}_{1},1\right)-E\left(\vb{x}-\vu{e}_{2},2\right)\right)\theta\left(\vb{x}\right)}
  \end{aligned}
\end{equation}
-- where, by expressing it as a vertex-dependent transformation, we
changed the point of view: instead of transforming the gauge field
with respect to the matter configuration, we look at it now as a transformation
that changes the matter field on each vertex, controlled by the electric
fields on the links around it. It changes the matter charge at each
vertex exactly by the amount of electric field divergence, and thanks
to the Gauss law it means that it is reduced to zero. This is the
controlled operation we wanted for decoupling the matter (while eliminating
the gauge constraints)! If we act with it on a gauge invariant state
in this Hilbert space, we will get a product of a trivial matter state
with a gauge field superposition that still contains the same physical
information, but without the symmetry: for an arbitrary gauge invariant
state,
\begin{equation}
\left|\psi\right\rangle =\underset{\left\{ Q\left(\vb{x}\right)\right\} }{\sum}\alpha\left(\left\{ Q\left(\vb{x}\right)\right\} \right)\left|E\left(\left\{ Q\left(\vb{x}\right)\right\} \right)\right\rangle _{\mathrm{Gauge}}\left|\left\{ Q\left(\vb{x}\right)\right\} \right\rangle _{\mathrm{Matter}}
\end{equation}
with a Gauss law at each vertex,
we have that
\begin{equation}
\mathcal{U}_{H}\left|\psi\right\rangle =\left|\left\{ Q\left(\vb{x}\right)=0\right\} \right\rangle _{\mathrm{Matter}}\otimes\underset{\left\{ Q\left(\vb{x}\right)\right\} }{\sum}\alpha\left(\left\{ Q\left(\vb{x}\right)\right\} \right)\left|E\left(\left\{ Q\left(\vb{x}\right)\right\} \right)\right\rangle _{\mathrm{Gauge}}
\end{equation}
without any Gauss law constraints.

If we consider the case of a $\mathbb{Z}_{2}$ gauge theory with such
Higgs matter, where both the gauge field and the matter are described
by spin half Hilbert spaces, $\mathcal{U}_{H}$, is nothing but a
controlled not (CNOT) operation, as discussed in~\cite{haegeman_gauging_2015}.

\subsection{Eliminating Fermionic Matter}

Next we wish to ask -- is there an analogy for fermionic matter? The
answer is: yes, but carefully:
 fermions come with their
special statistics and its complications, and thus one has to be extra cautious,
and act slightly differently than what we did above. Following~\cite{zohar_eliminating_2018}, we will review (and demonstrate for $U(1)$) how to replace the fermionic degrees of freedom by spins (hard-core bosons).

We go back now to the $U\left(1\right)$ lattice gauge theory with
fermionic matter discussed in the very beginning, with interaction
terms of the form
\begin{equation}
\widetilde{H}=\epsilon\underset{\vb{x},i=1,2}{\sum}\left(\psi^{\dagger}\left(\vb{x}\right)U\left(\vb{x},i\right)\psi\left(\vb{x}+\vu{e}_{i}\right)+h.c.\right).
\end{equation}

\begin{figure}[h]
  \centering
  \includegraphics{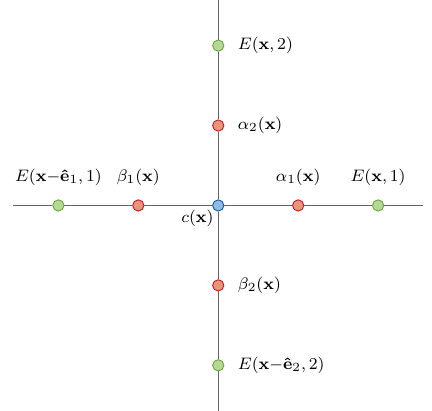} 
  \caption{
    Visual representation of the construction detailed in~\eqref{eq:el-ferm_def_c_eta} to~\eqref{eq:el-ferm_comm_eta}.
    For clarity, we show only the auxiliary modes that are newly introduced and the electric fields.
    On each vertex site $\vb{x}$ an auxiliary fermionic Majorana mode $c(\vb{x})$ (blue circle) is defined.
    The auxiliary modes $\alpha_i(\vb{x})$ and $\beta_i(\vb{x})$ (red circles) are introduced below and used in equation~\eqref{eq:el-ferm_def_f}.
    The electric fields on the respective links are shown as green circles.
  }
  \label{fig:el-ferm_def_aux_modes}
\end{figure}

In an (incomplete, as we shall see) analogy to the Higgs case, we
express the fermionic mode operators as
\begin{equation}
\psi^{\dagger}\left(\vb{x}\right)=\eta^{\dagger}\left(\vb{x}\right)c\left(\vb{x}\right),
\label{eq:el-ferm_def_c_eta}
\end{equation}
where $c\left(\vb{x}\right)$ is a fermionic Majorana mode, satisfying
the Clifford algebra
\begin{equation}
\left\{ c\left(\vb{x}\right),c\left(\vb{y}\right)\right\} =2\delta\left(\vb{x},\vb{y}\right)
\end{equation}
and the modes $\eta^{\dagger}\left(\vb{x}\right)$ have an on-site
fermionic anticommutation relation,
\begin{equation}
\left\{ \eta^{\dagger}\left(\vb{x}\right),\eta\left(\vb{x}\right)\right\} =1\quad;\left\{ \eta^{\dagger}\left(\vb{x}\right),c\left(\vb{x}\right)\right\} =0
\end{equation}
and off-site bosonic commutation relations,
\begin{equation}
\left[\eta^{\dagger}\left(\vb{x}\right),\eta\left(\vb{y}\right)\right]=\left[\eta\left(\vb{x}\right),\eta\left(\vb{y}\right)\right]=\left[c\left(\vb{x}\right),\eta\left(\vb{y}\right)\right]=0\quad\mathrm{for}\:\vb{x}\neq\vb{y}.
\label{eq:el-ferm_comm_eta}
\end{equation}
Figure~\ref{fig:el-ferm_def_aux_modes} shows an overview of all modes present on a vertex and its links.

The operator $\eta^{\dagger}\left(\vb{x}\right)$ -- almost analogous
to the radial field $R\left(\vb{x}\right)$ -- behaves like the
creation operator of a hard-core boson; if we can remove the operators
$c\left(\vb{x}\right)$ from the Hamiltonian, we can eventually
represent $\eta^{\dagger}\left(\vb{x}\right)$ by spin raising
operators, $\sigma_{+}\left(\vb{x}\right)$. In this sense, the
operators $c\left(\vb{x}\right)$ are analogous to $e^{i\theta\left(\vb{x}\right)}$
in the bosonic, Higgs case; indeed, one can think about the Majorana
mode operator obeying $c^{2}\left(\vb{x}\right)=1$ as a ``$\mathbb{Z}_{2}$
phase operator'' having to do with the fermionic $\mathbb{Z}_{2}$
symmetry. The analogy is not complete, though; in the Higgs case,
the radial and angular components were independent, commuting on-site
as well, but here, although commuting for different sites, on-site
we have $\left\{ \eta^{\dagger}\left(\vb{x}\right),c\left(\vb{x}\right)\right\} =0$:
the ``radial'' and ``angular'' modes are related through a $\mathbb{Z}_{2}$
equation. In fact, unlike in the Higgs case, the field $\eta^{\dagger}\left(\vb{x}\right)$
contains all the physical information, in the sense that its charge
participates in the $U\left(1\right)$ Gauss laws, while $c\left(\vb{x}\right)$
is only related to a $\mathbb{Z}_{2}$ symmetry -- the fermionic statistics.
These two fields are related through the local $\mathbb{Z}_{2}$ symmetry
manifested by $\left\{ \eta^{\dagger}\left(\vb{x}\right),c\left(\vb{x}\right)\right\} =0$.
Therefore, in the fermionic case the ``separation'', which is not
absolute, between ``radius'' and ``angle'' could be seen as separating
physics and statistics -- held together by the anticommutation relation.

In these terms, the interaction Hamiltonian has the form
\begin{equation}
  \widetilde{H}=\epsilon\underset{\vb{x},i=1,2}{\sum}\left(\eta^{\dagger}\left(\vb{x}\right)c\left(\vb{x}\right)U\left(\vb{x},i\right)c\left(\vb{x}+\vu{e}_{i}\right)\eta\left(\vb{x}+\vu{e}_{i}\right)+h.c.\right)
\end{equation}
Ideally, we would like to have a transformation, for which
\begin{equation}
U\left(\vb{x},i\right)\longrightarrow c\left(\vb{x}\right)U\left(\vb{x},i\right)c\left(\vb{x}+\vu{e}_{i}\right)
\end{equation}
and we are done. This, however, would be a problem: each link is transformed
from the left by $c\left(\vb{x}\right)$ and from the right by
$c\left(\vb{x}+\vu{e}_{i}\right)$. This is realized
with local transformations of the form
\begin{equation}
\mathcal{U}'_{F}\left(\vb{x}\right)=c^{G\left(\vb{x}\right)}=c^{\left(E\left(\vb{x},1\right)+E\left(\vb{x},2\right)-E\left(\vb{x}-\vu{e}_{1},1\right)-E\left(\vb{x}-\vu{e}_{2},2\right)\right)}.
\end{equation}
It works, thanks to the $\mathbb{Z}_{2}$ relation
\begin{equation}
e^{i\pi E}Ue^{-i\pi E}=-U
\end{equation}
responsible for
\begin{equation}
c^{E}Uc^{E}=-cU.
\end{equation}
However, these transformations, that ``rotate $U\left(\vb{x},i\right)$
by a fermionic phase'', do not have a fixed fermionic parity and
violate the fermionic superselection locally. It can be shown that
the product of all these, which is what we need, preserves fermionic
parity globally, but as it is locally broken, one has to define some
order for the product, that will eventually give rise to nonlocal
results.

\begin{figure}[h]
  \centering
  \includegraphics[]{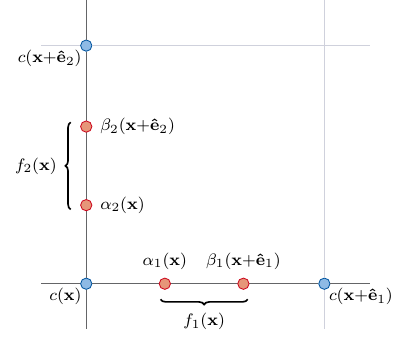} 
  \caption{
    Pictorial representation of~\eqref{eq:el-ferm_def_f}.
    The modes $f_i(\vb{x})$ related to vertex $\vb{x}$ in direction $i$ are formed by combining the modes $\alpha$ and $\beta$ on the respective link.
    The auxiliary modes $c(\vb{x})$ are depicted in blue as in Figure~\ref{fig:el-ferm_def_aux_modes}.
  }
  \label{fig:el-ferm_def_f}
\end{figure}

What we do, instead, is to introduce on each $\ell=\left(\vb{x},i\right)$
link two Majorana modes, $\alpha_{i}\left(\vb{x}\right)$ associated
with its beginning and $\beta_{i}\left(\vb{x}+\vu{e}_{i}\right)$
with its end. These operators do not appear in the original Hamiltonian
at all and hence we can fix the initial state of the respective auxiliary degrees of freedom as we wish. If the
original Hilbert space, without these modes, is $\mathcal{H}$, we
simply embed it within $\mathcal{H}\times\Omega$, where $\Omega$
is the one dimensional Hilbert space consisting of the Fock vacuum
annihilated by the link annihilation operators
\begin{equation}
  f_{i}\left(\vb{x}\right)=\frac{1}{2}\left(\alpha_{i}\left(\vb{x}\right)-i\beta_{i}\left(\vb{x}+\vu{e}_{i}\right)\right)
  \label{eq:el-ferm_def_f}
\end{equation}
as well as the vertex ones, $\chi\left(\vb{x}\right)$, for which
\begin{equation}
  c\left(\vb{x}\right)=\chi\left(\vb{x}\right)+\chi^{\dagger}\left(\vb{x}\right).
\end{equation}
The visual representation of equation~\eqref{eq:el-ferm_def_f} is shown in Figure~\ref{fig:el-ferm_def_f}.

We define the local transformation
\begin{equation}
\mathcal{U}_{F}\left(\vb{x}\right)=\left(ic\left(\vb{x}\right)\beta_{2}\left(\vb{x}\right)\right)^{E\left(\vb{x}-\vu{e}_{2},2\right)}\left(ic\left(\vb{x}\right)\beta_{1}\left(\vb{x}\right)\right)^{E\left(\vb{x}-\vu{e}_{1},1\right)}\left(ic\left(\vb{x}\right)\alpha_{2}\left(\vb{x}\right)\right)^{E\left(\vb{x},2\right)}\left(ic\left(\vb{x}\right)\alpha_{1}\left(\vb{x}\right)\right)^{E\left(\vb{x},1\right)}
\end{equation}
It has an even fermionic parity, and thus preserves the fermionic
superselection rule locally; thanks to that, and having different
fermionic modes at each vertex, $\left[\mathcal{U}_{F}\left(\vb{x}\right),\mathcal{U}_{F}\left(\vb{x}'\right)\right]=0$,
the transformation
\begin{equation}
  \mathcal{U}_{F}=\underset{\vb{x}}{\prod}\mathcal{U}_{F}\left(\vb{x}\right)
\end{equation}
is well defined, since no ordering has to be specified for the product.
We obtain that
\begin{equation}
\mathcal{U}_{F}\widetilde{H}\mathcal{U}_{F}^{\dagger}=\epsilon\underset{\vb{x},i=1,2}{\sum}\left(\xi_{i}\eta^{\dagger}\left(\vb{x}\right)\alpha_{i}\left(\vb{x}\right)U\left(\vb{x},i\right)\beta_{i}\left(\vb{x}+\vu{e}_{i}\right)\eta\left(\vb{x}+\vu{e}_{i}\right)+h.c.\right),
\end{equation}
where $\xi_{i}$ are phases (signs) that depend on the electric fields
on some links around the link's edges~\cite{zohar_eliminating_2018}. Since the
initial state for the auxiliary link fermions was the vacuum, and
one may check that $\left[\mathcal{U}_{F},\alpha_{i}\left(\vb{x}\right)\beta_{i}\left(\vb{x}+\vu{e}_{i}\right)\right]=0$,
they remain so after the transformation, and could be replaced by $-i$. 
The hard-core bosonic operators $\eta^{\dagger}\left(\vb{x}\right)$
can be effectively replaced by spin raising operators $\sigma_{+}\left(\vb{x}\right)$
(in a process that involves local, string-less Jordan Wigner transformations
-- see~\cite{zohar_eliminating_2018}), and we obtain a transformed interaction
Hamiltonian of the form
%FIXME: Add H_?
\begin{equation}
-i\epsilon\underset{\vb{x},i=1,2}{\sum}\left(\xi_{i}\sigma_{+}\left(\vb{x}\right)U\left(\vb{x},i\right)\sigma_{-}\left(\vb{x}+\vu{e}_{i}\right)+h.c.\right).
\end{equation}
Other terms in the Hamiltonian also transform in a way that leaves
no fermions present~\cite{zohar_eliminating_2018}. Therefore we see that our
lattice gauge theory with fermionic matter was mapped to one with
hard core bosonic matter (that can be now removed straightforwardly
using the Gauss laws). The lattice rotations are broken by the $\xi_{i}$
operators, whose role is to keep the right statistics and commutation
relations~\cite{zohar_eliminating_2018}.

In order to demonstrate it simply, let us look at the one dimensional
case, where
\begin{equation}
\mathcal{U}_{F}\left(x\right)=\left(ic\left(x\right)\beta\left(x\right)\right)^{E\left(x-1\right)}\left(ic\left(x\right)\alpha\left(x\right)\right)^{E\left(x\right)}
\end{equation}
and one simply obtains
\begin{equation}
-i\epsilon\underset{x}{\sum}\left(e^{i\pi E\left(x-1\right)}\sigma_{+}\left(x\right)U\left(x\right)\sigma_{-}\left(x\right)+h.c.\right).
\end{equation}

As the "radial" and "angular" degrees of freedom were related
to each other, the symmetry is not broken, and a matter field still
exists; however, it is not fermionic. Unlike in the Higgs case, we did not break the continuous gauge symmetry completely
and did not eliminate the matter: we only exploited the $\mathbb{Z}_2$ subgroup of the gauge group, $U(1)$ in our example,
to eliminate the fermionic nature of the matter.
In some cases, as the one we discuss, one may move on and eliminate the hard-core bosons using
a simple, non-fermionic controlled operation as the one shown above
for Higgs field; the difference will be, that now the bosons are hard-core,
so projectors for the right values of electric field divergence must
be introduced.

In fact, this can be done for every lattice gauge
theory whose gauge group includes a $\mathbb{Z}_{2}$ normal subgroup~\cite{zohar_eliminating_2018} 
-- for example, $\mathbb{Z}_{2N},U\left(N\right),SU\left(2N\right)$:
the important property is to have an operator $E$, for which the
$\mathbb{Z}_{2}$ relation
\begin{equation}
e^{i\pi E}Ue^{-i\pi E}=-U
\end{equation}
holds. This is the basis for our transformation. Groups with a $\mathbb{Z}_{2}$
normal subgroup satisfy that. For other cases, e.g. $SU\left(2N+1\right),$
one can introduce an auxiliary $\mathbb{Z}_{2}$ gauge field -- that
is, extend the gauge symmetry to e.g. $SU\left(2N+1\right)\times\mathbb{Z}_{2}$,
and use it for the elimination of fermions (and their replacement by hard core bosons) in a process as the one
discussed above. It can be done in a way that preserves the physical
properties of the original model (no dynamics is added for the auxiliary
$\mathbb{Z}_{2}$ gauge field). However, if one has more than one spinor component on each site,
extra caution has to be taken when representing the hard core bosons by spins: \emph{local}
Jordan-Wigner transformations must be used \cite{zohar_eliminating_2018}.

The relation to $\mathbb{Z}_{2}$ is not accidental: any "reasonable"
fermionic theory without supersymmetry has a global $\mathbb{Z}_{2}$ symmetry -- parity
superselection, which we have already discussed. When we have a $\mathbb{Z}_{2}$
local symmetry, we can construct a transformation as above and eliminate
the fermions locally since their information is ``saved'' by the
gauge field. If there is no such field we can add it with some minimal
coupling procedure and then do as above: so, in order to remove the
fermionic statistics, one simply has to gauge the global $\mathbb{Z}_{2}$
symmetry associated with it and make it local.

When the gauge group also has a $U(1)$ normal subgroup, as in the case of $U(N)$, one can go one step further and completely eliminate the matter as in the Higgs unitary gauge \cite{fradkin_phase_1979}, even in the cases of hard-core bosonic or fermionic matter~\cite{zohar_removing_2019}.
As a result of this procedure, only gauge field degrees of freedom will appear, but the symmetry will be broken as in the original case;
however, since the matter degrees of freedom were bounded tobegin with, some local projectors will be added, as well as local constraints - which do not appear in the Abelian Higgs case~\cite{fradkin_phase_1979}. 
Since it is done unitarily, the inverse procedure can be carried out as well, in which one starts with a theory that contains gauge-like fields on the links (without gauge symmetry), and unitarily couples them minimally to matter. 
It is another manifestation of the fact that the Gauss law is explicitly solved for the matter but cannot be solved for the gauge fields: as argued in the beginning, introducing gauge fields to globally invariant matter theories by minimal coupling cannot be done unitarily, but introducing matter fields that will be minimally coupled to vector fields is possible unitarily, using the inverse of the above procedure.
\section*{Acknowledgments}
The first version of these lecture notes was prepared for two lectures given by Erez Zohar in the Focus week \emph{Tensor Networks and Entanglement} of the workshop \emph{Entanglement in Quantum System}, at the Galileo Galilei Institute for Theoretical Physics (GGI), Florence, Italy in June 2018. The current version was prepared by both authors, including further information and details.

Patrick Emonts acknowledges support from the International Max-Planck Research School for Quantum Science and Technology (IMPRS-QST) as well as support by the EU-QUANTERA project QTFLAG (BMBF grant No. 13N14780).

\begin{appendix}
  \section{Tensor Network Notation\label{app:tn_notation}}
  This short introduction to tensor network notation is added in order to make these lecture notes self-contained and does replace not further reading~\cite{bridgeman_hand-waving_2017,orus_practical_2014, verstraete_matrix_2008}.
  For the ease of description, we will have a look at an arbitrary state of a spin system and rewrite it as a matrix product state (MPS).
  This description can be readily generalized to the description of two-dimensional systems with a PEPS description.

  We consider an arbitrary state of a spin system of $N$ spins -- say spin-$\frac{1}{2}$ and write it as a superposition of states in the z-basis
  \begin{equation}
    \ket{\Psi}=\sum_{\sigma_1\ldots\sigma_N}c_{\sigma_1,\sigma_2\ldots\sigma_N}\ket{\sigma_1\sigma_2 \ldots \sigma_N}.
    \label{eq:app_arb_state}
  \end{equation}
  In general, the coefficients $c$ depend on the configuration of all spins in the system.
  As an Ansatz, we can rewrite the coefficients $c_{\sigma_1,\sigma_2\ldots\sigma_N}$ as a product of matrices (hence the name, MPS)
  \begin{equation}
    \ket{\Psi}=\sum_{\sigma_1\ldots\sigma_N}\underbrace{\sum_{a_1,\ldots,a_{N-1}}A^{\sigma_1}_{1,a_1}A^{\sigma_2}_{a_1,a_2}\cdots A^{\sigma_{N-1}}_{a_{N-2},a_{N-1}}A^{\sigma_N}_{a_{N-1},1}}_{c_{\sigma_1,\sigma_2\ldots\sigma_N}}\ket{\sigma_1\sigma_2 \ldots \sigma_N}.
    \label{eq:app_def_mps}
  \end{equation}
  The objects $A$ are called \emph{tensors} and constitute the elementary building block of the Ansatz.
  We call them tensors since they are objects with more than two indices.
  We do not imply any transformation properties generally associated with the name.
  They carry two types of indices: \emph{physical indices} $\sigma$ and \emph{virtual indices} $a$.
  While the physical ones correspond to the indices of the coefficients in~\eqref{eq:app_arb_state}, the virtual indices have no physical meaning.
  They are added in order to define the contraction of the tensors.

\begin{figure}
    \centering
    \includegraphics{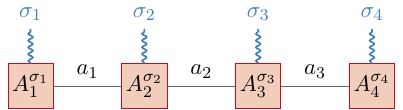}
    \caption{Pictorial representation of a matrix product state. 
      The corresponding mathematical description is detailed in~\eqref{eq:app_def_mps}.
      Physical indices/legs are drawn as wavy lines while virtual indices/legs are straight.}
    \label{fig:app_mps}
  \end{figure}

  Equation~\eqref{eq:app_def_mps} is depicted in Figure~\ref{fig:app_mps}.
  The squares correspond to the tensors $A$ and have three legs which correspond to the three indices.
  As a convention for these notes, we draw the physical indices $\sigma$ with wavy lines and the virtual legs as straight lines.
  If a leg connects two tensors, it will be contracted, i.e. the summation of this index will be executed.
  Since all virtual indices in the formulation of~\eqref{eq:app_def_mps} are summed over, they are all connected.
  
\begin{figure}
    \centering
    \includegraphics[scale=0.5]{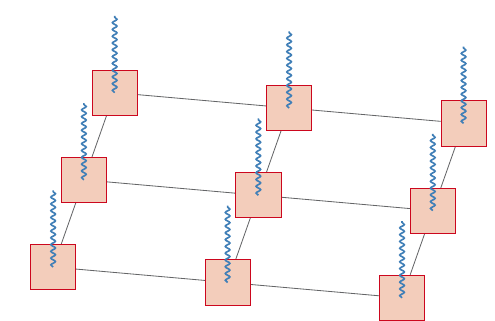}
    \caption{Pictorial representation of a PEPS. The virtual degrees of freedom connect tensors in the x- and y-direction since the system is three-dimensional.
    Physical indices/legs are drawn as wavy lines while virtual indices/legs are straight.}
    \label{fig:app_peps}
  \end{figure}

  If the system of interest has more than one spatial dimension, we can adapt our description by adding more virtual indices and describe the system with a two-dimensional PEPS.
  The pictorial representation of PEPS is shown in Figure~\ref{fig:app_peps}.
  As for the MPS case, virtual indices are depicted as straight legs and indices are contracted if their corresponding legs are connected.
\end{appendix}

\clearpage
\bibliographystyle{SciPost_bibstyle}
\bibliography{ref}

\begin{thebibliography}{10}
\providecommand{\url}[1]{\texttt{#1}}
\providecommand{\urlprefix}{URL }
\expandafter\ifx\csname urlstyle\endcsname\relax
  \providecommand{\doi}[1]{doi:\discretionary{}{}{}#1}\else
  \providecommand{\doi}{doi:\discretionary{}{}{}\begingroup
  \urlstyle{rm}\Url}\fi
\providecommand{\eprint}[2][]{\url{#2}}

\bibitem{wilson_confinement_1974}
K.~G. Wilson,
\newblock \emph{Confinement of quarks},
\newblock Physical Review D \textbf{10}(8), 2445 (1974),
\newblock \doi{10.1103/PhysRevD.10.2445}.

\bibitem{gross_ultraviolet_1973}
D.~J. Gross and F.~Wilczek,
\newblock \emph{Ultraviolet {{Behavior}} of {{Non}}-{{Abelian Gauge
  Theories}}},
\newblock Physical Review Letters \textbf{30}(26), 1343 (1973),
\newblock \doi{10.1103/PhysRevLett.30.1343}.

\bibitem{kogut_introduction_1979}
J.~B. Kogut,
\newblock \emph{An introduction to lattice gauge theory and spin systems},
\newblock Reviews of Modern Physics \textbf{51}(4), 659 (1979),
\newblock \doi{10.1103/RevModPhys.51.659}.

\bibitem{flag_working_group_review_2014}
{FLAG Working Group}, S.~Aoki, Y.~Aoki, C.~Bernard, T.~Blum, G.~Colangelo,
  M.~Della~Morte, S.~D{\"u}rr, A.~X. {El-Khadra}, H.~Fukaya, R.~Horsley,
  A.~J{\"u}ttner \emph{et~al.},
\newblock \emph{Review of lattice results concerning low-energy particle
  physics},
\newblock The European Physical Journal C \textbf{74}(9) (2014),
\newblock \doi{10.1140/epjc/s10052-014-2890-7}.

\bibitem{troyer_computational_2005}
M.~Troyer and U.-J. Wiese,
\newblock \emph{Computational {{Complexity}} and {{Fundamental Limitations}} to
  {{Fermionic Quantum Monte Carlo Simulations}}},
\newblock Physical Review Letters \textbf{94}(17) (2005),
\newblock \doi{10.1103/PhysRevLett.94.170201}.

\bibitem{fukushima_phase_2011}
K.~Fukushima and T.~Hatsuda,
\newblock \emph{The phase diagram of dense {{QCD}}},
\newblock Reports on Progress in Physics \textbf{74}(1), 014001 (2011),
\newblock \doi{10.1088/0034-4885/74/1/014001}.

\bibitem{zohar_quantum_2016}
E.~Zohar, J.~I. Cirac and B.~Reznik,
\newblock \emph{Quantum simulations of lattice gauge theories using ultracold
  atoms in optical lattices},
\newblock Reports on Progress in Physics \textbf{79}(1), 014401 (2016),
\newblock \doi{10.1088/0034-4885/79/1/014401}.

\bibitem{wiese_ultracold_2013}
U.-J. Wiese,
\newblock \emph{Ultracold quantum gases and lattice systems: Quantum simulation
  of lattice gauge theories},
\newblock Annalen der Physik \textbf{525}(10-11), 777 (2013),
\newblock \doi{10.1002/andp.201300104}.

\bibitem{dalmonte_lattice_2016}
M.~Dalmonte and S.~Montangero,
\newblock \emph{Lattice gauge theory simulations in the quantum information
  era},
\newblock Contemporary Physics \textbf{57}(3), 388 (2016),
\newblock \doi{10.1080/00107514.2016.1151199}.

\bibitem{verstraete_matrix_2008}
F.~Verstraete, V.~Murg and J.~Cirac,
\newblock \emph{Matrix product states, projected entangled pair states, and
  variational renormalization group methods for quantum spin systems},
\newblock Advances in Physics \textbf{57}(2), 143 (2008),
\newblock \doi{10.1080/14789940801912366}.

\bibitem{orus_practical_2014}
R.~Or{\'u}s,
\newblock \emph{A practical introduction to tensor networks: {{Matrix}} product
  states and projected entangled pair states},
\newblock Annals of Physics \textbf{349}, 117 (2014),
\newblock \doi{10.1016/j.aop.2014.06.013}.

\bibitem{kogut_hamiltonian_1975}
J.~Kogut and L.~Susskind,
\newblock \emph{Hamiltonian formulation of {{Wilson}}'s lattice gauge
  theories},
\newblock Physical Review D \textbf{11}(2), 395 (1975),
\newblock \doi{10.1103/PhysRevD.11.395}.

\bibitem{zohar_fermionic_2015}
E.~Zohar, M.~Burrello, T.~B. Wahl and J.~I. Cirac,
\newblock \emph{Fermionic projected entangled pair states and local {{U}}(1)
  gauge theories},
\newblock Annals of Physics \textbf{363}, 385 (2015),
\newblock \doi{10.1016/j.aop.2015.10.009}.

\bibitem{zohar_building_2016}
E.~Zohar and M.~Burrello,
\newblock \emph{Building projected entangled pair states with a local gauge
  symmetry},
\newblock New Journal of Physics \textbf{18}(4), 043008 (2016),
\newblock \doi{10.1088/1367-2630/18/4/043008}.

\bibitem{zohar_projected_2016}
E.~Zohar, T.~B. Wahl, M.~Burrello and J.~I. Cirac,
\newblock \emph{Projected {{Entangled Pair States}} with non-{{Abelian}} gauge
  symmetries: {{An SU}}(2) study},
\newblock Annals of Physics \textbf{374}, 84 (2016),
\newblock \doi{10.1016/j.aop.2016.08.008}.

\bibitem{zohar_digital_2017}
E.~Zohar, A.~Farace, B.~Reznik and J.~I. Cirac,
\newblock \emph{Digital {{Quantum Simulation}} of {{Z}} 2 {{Lattice Gauge
  Theories}} with {{Dynamical Fermionic Matter}}},
\newblock Physical Review Letters \textbf{118}(7) (2017),
\newblock \doi{10.1103/PhysRevLett.118.070501}.

\bibitem{zohar_digital_2017-1}
E.~Zohar, A.~Farace, B.~Reznik and J.~I. Cirac,
\newblock \emph{Digital lattice gauge theories},
\newblock Physical Review A \textbf{95}(2) (2017),
\newblock \doi{10.1103/PhysRevA.95.023604}.

\bibitem{zohar_combining_2018}
E.~Zohar and J.~I. Cirac,
\newblock \emph{Combining tensor networks with {{Monte Carlo}} methods for
  lattice gauge theories},
\newblock Physical Review D \textbf{97}(3) (2018),
\newblock \doi{10.1103/PhysRevD.97.034510}.

\bibitem{zohar_eliminating_2018}
E.~Zohar and J.~I. Cirac,
\newblock \emph{Eliminating fermionic matter fields in lattice gauge theories},
\newblock Physical Review B \textbf{98}(7) (2018),
\newblock \doi{10.1103/PhysRevB.98.075119}.

\bibitem{zohar_removing_2019}
E.~Zohar and J.~I. Cirac,
\newblock \emph{Removing staggered fermionic matter in {$U(N)$} and {$SU(N)$}
  lattice gauge theories},
\newblock Physical Review D \textbf{99}(11) (2019),
\newblock \doi{10.1103/PhysRevD.99.114511}.

\bibitem{peskin_introduction_1995}
M.~E. Peskin and D.~V. Schroeder,
\newblock \emph{An Introduction to Quantum Field Theory},
\newblock {Addison-Wesley Pub. Co}, {Reading, Mass},
\newblock ISBN 978-0-201-50397-5 (1995).

\bibitem{susskind_lattice_1977}
L.~Susskind,
\newblock \emph{Lattice fermions},
\newblock Physical Review D \textbf{16}(10), 3031 (1977),
\newblock \doi{10.1103/PhysRevD.16.3031}.

\bibitem{zohar_formulation_2015}
E.~Zohar and M.~Burrello,
\newblock \emph{Formulation of lattice gauge theories for quantum simulations},
\newblock Physical Review D \textbf{91}(5) (2015),
\newblock \doi{10.1103/PhysRevD.91.054506}.

\bibitem{savit_duality_1980}
R.~Savit,
\newblock \emph{Duality in field theory and statistical systems},
\newblock Reviews of Modern Physics \textbf{52}(2), 453 (1980),
\newblock \doi{10.1103/RevModPhys.52.453}.

\bibitem{buividovich_entanglement_2008}
P.~Buividovich and M.~Polikarpov,
\newblock \emph{Entanglement entropy in gauge theories and the holographic
  principle for electric strings},
\newblock Physics Letters B \textbf{670}(2), 141 (2008),
\newblock \doi{10.1016/j.physletb.2008.10.032}.

\bibitem{tagliacozzo_entanglement_2011}
L.~Tagliacozzo and G.~Vidal,
\newblock \emph{Entanglement renormalization and gauge symmetry},
\newblock Physical Review B \textbf{83}(11), 115127 (2011),
\newblock \doi{10.1103/PhysRevB.83.115127}.

\bibitem{banuls_mass_2013}
M.~Ba{\~n}uls, K.~Cichy, J.~Cirac and K.~Jansen,
\newblock \emph{The mass spectrum of the {{Schwinger}} model with matrix
  product states},
\newblock Journal of High Energy Physics \textbf{2013}(11) (2013),
\newblock \doi{10.1007/JHEP11(2013)158}.

\bibitem{martinez_real-time_2016}
E.~A. Martinez, C.~A. Muschik, P.~Schindler, D.~Nigg, A.~Erhard, M.~Heyl,
  P.~Hauke, M.~Dalmonte, T.~Monz, P.~Zoller and R.~Blatt,
\newblock \emph{Real-time dynamics of lattice gauge theories with a few-qubit
  quantum computer},
\newblock Nature \textbf{534}, 516 (2016),
\newblock \doi{10.1038/nature18318}.

\bibitem{suzuki_decomposition_1985}
M.~Suzuki,
\newblock \emph{Decomposition formulas of exponential operators and {{Lie}}
  exponentials with some applications to quantum mechanics and statistical
  physics},
\newblock Journal of Mathematical Physics \textbf{26}(4), 601 (1985),
\newblock \doi{10.1063/1.526596}.

\bibitem{lloyd_universal_1996}
S.~Lloyd,
\newblock \emph{Universal {{Quantum Simulators}}},
\newblock Science \textbf{273}(5278), 1073 (1996),
\newblock \doi{10.1126/science.273.5278.1073}.

\bibitem{bender_digital_2018}
J.~Bender, E.~Zohar, A.~Farace and J.~I. Cirac,
\newblock \emph{Digital quantum simulation of lattice gauge theories in three
  spatial dimensions},
\newblock New Journal of Physics \textbf{20}(9), 093001 (2018),
\newblock \doi{10.1088/1367-2630/aadb71}.

\bibitem{bridgeman_hand-waving_2017}
J.~C. Bridgeman and C.~T. Chubb,
\newblock \emph{Hand-waving and {{Interpretive Dance}}: {{An Introductory
  Course}} on {{Tensor Networks}}},
\newblock Journal of Physics A: Mathematical and Theoretical \textbf{50}(22),
  223001 (2017),
\newblock \doi{10.1088/1751-8121/aa6dc3},
\newblock \eprint{1603.03039}.

\bibitem{haegeman_gauging_2015}
J.~Haegeman, K.~Van~Acoleyen, N.~Schuch, J.~I. Cirac and F.~Verstraete,
\newblock \emph{Gauging {{Quantum States}}: {{From Global}} to {{Local
  Symmetries}} in {{Many}}-{{Body Systems}}},
\newblock Phys. Rev. X \textbf{5}(1), 011024 (2015),
\newblock \doi{10.1103/PhysRevX.5.011024}.

\bibitem{tagliacozzo_tensor_2014}
L.~Tagliacozzo, A.~Celi and M.~Lewenstein,
\newblock \emph{Tensor {{Networks}} for {{Lattice Gauge Theories}} with
  {{Continuous Groups}}},
\newblock Phys. Rev. X \textbf{4}(4), 041024 (2014),
\newblock \doi{10.1103/PhysRevX.4.041024}.

\bibitem{kraus_fermionic_2010}
C.~V. Kraus, N.~Schuch, F.~Verstraete and J.~I. Cirac,
\newblock \emph{Fermionic projected entangled pair states},
\newblock Physical Review A \textbf{81}(5) (2010),
\newblock \doi{10.1103/PhysRevA.81.052338}.

\bibitem{bravyi_lagrangian_2005}
S.~Bravyi,
\newblock \emph{Lagrangian representation for fermionic linear optics},
\newblock Quantum Inf. and Comp. \textbf{5}(3), 216 (2005).

\bibitem{yang_chiral_2015}
S.~Yang, T.~B. Wahl, H.-H. Tu, N.~Schuch and J.~I. Cirac,
\newblock \emph{Chiral {{Projected Entangled}}-{{Pair State}} with
  {{Topological Order}}},
\newblock Phys. Rev. Lett. \textbf{114}(10), 106803 (2015),
\newblock \doi{10.1103/PhysRevLett.114.106803}.

\bibitem{fradkin_phase_1979}
E.~Fradkin and S.~H. Shenker,
\newblock \emph{Phase diagrams of lattice gauge theories with {{Higgs}}
  fields},
\newblock Phys. Rev. D \textbf{19}(12), 3682 (1979),
\newblock \doi{10.1103/PhysRevD.19.3682}.

\bibitem{white_density_1992}
S.~R. White,
\newblock \emph{Density matrix formulation for quantum renormalization groups},
\newblock Physical Review Letters \textbf{69}(19), 2863 (1992),
\newblock \doi{10.1103/PhysRevLett.69.2863}.

\bibitem{byrnes_density_2002}
T.~M.~R. Byrnes, P.~Sriganesh, R.~J. Bursill and C.~J. Hamer,
\newblock \emph{Density matrix renormalization group approach to the massive
  {{Schwinger}} model},
\newblock Phys. Rev. D \textbf{66}(1), 013002 (2002),
\newblock \doi{10.1103/PhysRevD.66.013002}.

\bibitem{sugihara_matrix_2005}
T.~Sugihara,
\newblock \emph{Matrix product representation of gauge invariant states in a
  $\mathbb{Z}_2$ lattice gauge theory},
\newblock Journal of High Energy Physics \textbf{2005}(07), 022 (2005),
\newblock \doi{10.1088/1126-6708/2005/07/022}.

\bibitem{buyens_matrix_2014}
B.~Buyens, J.~Haegeman, K.~Van~Acoleyen, H.~Verschelde and F.~Verstraete,
\newblock \emph{Matrix {{Product States}} for {{Gauge Field Theories}}},
\newblock Physical Review Letters \textbf{113}(9) (2014),
\newblock \doi{10.1103/PhysRevLett.113.091601}.

\bibitem{silvi_lattice_2014}
P.~Silvi, E.~Rico, T.~Calarco and S.~Montangero,
\newblock \emph{Lattice gauge tensor networks},
\newblock New Journal of Physics \textbf{16}(10), 103015 (2014),
\newblock \doi{10.1088/1367-2630/16/10/103015}.

\bibitem{rico_tensor_2014}
E.~Rico, T.~Pichler, M.~Dalmonte, P.~Zoller and S.~Montangero,
\newblock \emph{Tensor {{Networks}} for {{Lattice Gauge Theories}} and {{Atomic
  Quantum Simulation}}},
\newblock Physical Review Letters \textbf{112}(20) (2014),
\newblock \doi{10.1103/PhysRevLett.112.201601}.

\bibitem{saito_temperature_2014}
H.~Saito, M.~C. Ba{\~n}uls, K.~Cichy, J.~I. Cirac and K.~Jansen,
\newblock \emph{The temperature dependence of the chiral condensate in the
  {{Schwinger}} model with {{Matrix Product States}}},
\newblock arXiv:1412.0596; PoS(LATTICE2014)302  (2014).

\bibitem{kuhn_non-abelian_2015}
S.~K{\"u}hn, E.~Zohar, J.~I. Cirac and M.~C. Ba{\~n}uls,
\newblock \emph{Non-{{Abelian}} string breaking phenomena with matrix product
  states},
\newblock Journal of High Energy Physics \textbf{2015}(7), 1 (2015),
\newblock \doi{10.1007/JHEP07(2015)130}.

\bibitem{banuls_thermal_2015}
M.~C. Ba{\~n}uls, K.~Cichy, J.~I. Cirac, K.~Jansen and H.~Saito,
\newblock \emph{Thermal evolution of the {{Schwinger}} model with matrix
  product operators},
\newblock Physical Review D \textbf{92}(3) (2015),
\newblock \doi{10.1103/PhysRevD.92.034519}.

\bibitem{banuls_chiral_2016}
M.~C. Ba{\~n}uls, K.~Cichy, K.~Jansen and H.~Saito,
\newblock \emph{Chiral condensate in the {{Schwinger}} model with matrix
  product operators},
\newblock Phys. Rev. D \textbf{93}(9), 094512 (2016),
\newblock \doi{10.1103/PhysRevD.93.094512}.

\bibitem{pichler_real-time_2016}
T.~Pichler, M.~Dalmonte, E.~Rico, P.~Zoller and S.~Montangero,
\newblock \emph{Real-{{Time Dynamics}} in {{U}}(1) {{Lattice Gauge Theories}}
  with {{Tensor Networks}}},
\newblock Phys. Rev. X \textbf{6}(1), 011023 (2016),
\newblock \doi{10.1103/PhysRevX.6.011023}.

\bibitem{silvi_finite-density_2017}
P.~Silvi, E.~Rico, M.~Dalmonte, F.~Tschirsich and S.~Montangero,
\newblock \emph{Finite-density phase diagram of a ( 1 + 1 ) - d non-abelian
  lattice gauge theory with tensor networks},
\newblock Quantum \textbf{1}, 9 (2017),
\newblock \doi{10.22331/q-2017-04-25-9}.

\bibitem{milsted_matrix_2016}
A.~Milsted,
\newblock \emph{Matrix product states and the non-{{Abelian}} rotor model},
\newblock Phys. Rev. D \textbf{93}(8), 085012 (2016),
\newblock \doi{10.1103/PhysRevD.93.085012}.

\bibitem{buyens_hamiltonian_2016}
B.~Buyens, F.~Verstraete and K.~Van~Acoleyen,
\newblock \emph{Hamiltonian simulation of the {{Schwinger}} model at finite
  temperature},
\newblock Physical Review D \textbf{94}(8) (2016),
\newblock \doi{10.1103/PhysRevD.94.085018}.

\bibitem{banuls_multi-flavor_2016}
M.~Ba{\~n}uls, K.~Cichy, J.~Cirac, K.~Jansen, S.~K{\"u}hn and H.~Saito,
\newblock \emph{The multi-flavor {{Schwinger}} model with chemical potential -
  {{Overcoming}} the sign problem with {{Matrix Product States}}},
\newblock arXiv:1611.01458 [hep-lat]  (2016).

\bibitem{buyens_real-time_2017}
B.~Buyens, J.~Haegeman, F.~Hebenstreit, F.~Verstraete and K.~Van~Acoleyen,
\newblock \emph{Real-time simulation of the {{Schwinger}} effect with matrix
  product states},
\newblock Physical Review D \textbf{96}(11) (2017),
\newblock \doi{10.1103/PhysRevD.96.114501}.

\bibitem{banuls_density_2017}
M.~C. Ba{\~n}uls, K.~Cichy, J.~I. Cirac, K.~Jansen and S.~K{\"u}hn,
\newblock \emph{Density {{Induced Phase Transitions}} in the {{Schwinger
  Model}}: {{A Study}} with {{Matrix Product States}}},
\newblock Phys. Rev. Lett. \textbf{118}(7), 071601 (2017),
\newblock \doi{10.1103/PhysRevLett.118.071601}.

\bibitem{banuls_mari_carmen_towards_2017}
{Ba{\~n}uls, Mari Carmen}, {Cichy, Krzysztof}, {Ignacio Cirac, J.}, {Jansen,
  Karl}, {K{\"u}hn, Stefan} and {Saito, Hana},
\newblock \emph{Towards overcoming the {{Monte Carlo}} sign problem with tensor
  networks},
\newblock EPJ Web Conf. \textbf{137}, 04001 (2017),
\newblock \doi{10.1051/epjconf/201713704001}.

\bibitem{banuls_efficient_2017}
M.~C. Ba{\~n}uls, K.~Cichy, J.~I. Cirac, K.~Jansen and S.~K{\"u}hn,
\newblock \emph{Efficient {{Basis Formulation}} for ( 1 + 1 )-{{Dimensional
  SU}}(2) {{Lattice Gauge Theory}}: {{Spectral Calculations}} with {{Matrix
  Product States}}},
\newblock Physical Review X \textbf{7}(4) (2017),
\newblock \doi{10.1103/PhysRevX.7.041046}.

\bibitem{carmen_banuls_tensor_2018}
M.~Carmen~Ba{\~n}uls, K.~Cichy, Y.-J. Kao, C.-J. David~Lin, Y.-P. Lin and
  D.~{Tao-Lin Tan},
\newblock \emph{Tensor {{Network}} study of the (1+1)-dimensional {{Thirring
  Model}}},
\newblock EPJ Web of Conferences \textbf{175}, 11017 (2018),
\newblock \doi{10.1051/epjconf/201817511017}.

\bibitem{silvi_tensor_2019}
P.~Silvi, F.~Tschirsich, M.~Gerster, J.~J{\"u}nemann, D.~Jaschke, M.~Rizzi and
  S.~Montangero,
\newblock \emph{The {{Tensor Networks Anthology}}: {{Simulation}} techniques
  for many-body quantum lattice systems},
\newblock SciPost Physics Lecture Notes  (2019),
\newblock \doi{10.21468/SciPostPhysLectNotes.8}.

\bibitem{banuls_tensor_2018}
M.~C. Ba{\~n}uls, K.~Cichy, J.~I. Cirac, K.~Jansen and S.~K{\"u}hn,
\newblock \emph{Tensor {{Networks}} and their use for {{Lattice Gauge
  Theories}}},
\newblock arXiv:1810.12838 [hep-lat]  (2018),
\newblock \eprint{1810.12838}.

\bibitem{silvi_tensor_2019-1}
P.~Silvi, Y.~Sauer, F.~Tschirsich and S.~Montangero,
\newblock \emph{Tensor {{Network Simulation}} of compact one-dimensional
  lattice {{Quantum Chromodynamics}} at finite density},
\newblock arXiv:1901.04403 [hep-lat, physics:quant-ph]  (2019),
\newblock \eprint{1901.04403}.

\bibitem{tschirsich_phase_2019}
F.~Tschirsich, S.~Montangero and M.~Dalmonte,
\newblock \emph{Phase diagram and conformal string excitations of square ice
  using gauge invariant matrix product states},
\newblock SciPost Physics \textbf{6}(3) (2019),
\newblock \doi{10.21468/SciPostPhys.6.3.028}.

\end{thebibliography}
%\nolinenumbers
\end{document}